\documentclass[12pt]{article}
\usepackage{amsmath}
\usepackage{amssymb}
\usepackage{graphicx}
\usepackage{xcolor}
\newcommand{\Tr}{{\rm Tr}}
\newcommand{\Z}{\mathbb{Z}}
\newcommand{\R}{\mathbb{R}}

\newcommand{\ket}{\rangle}
\newcommand{\de}{\partial}
\newcommand{\cP}{\mathcal{P}}
\newcommand{\cH}{\mathcal{H}}
\newcommand{\torus}[2]{#1\,\,\! \underset{\displaystyle #2}{
\begin{array}{|c|}
\hline \hspace{7pt} \\[3pt] 
\hline 
\end{array}
}}
\newcommand\blfootnote[1]{%
  \begingroup
  \renewcommand\thefootnote{}\footnote{#1}%
  \addtocounter{footnote}{-1}%
  \endgroup
}
\numberwithin{equation}{section}
\usepackage[unicode]{hyperref}
\hypersetup{colorlinks=true, breaklinks=true, linkcolor=black, menucolor=black, urlcolor=black!20!green!70!blue,citecolor=black!20!green!70!blue}

\begin{document}

\vspace*{-1.5cm}
\thispagestyle{empty}
\begin{flushright}
AEI-2012-087
\end{flushright}
\vspace*{2.5cm}
\begin{center}
{\Large
{\bf The geometry of the limit of $\boldsymbol{N=2}$ minimal models}}
\vspace{2.5cm}

{\large Stefan Fredenhagen, Cosimo Restuccia}
\blfootnote{{\tt E-mail: FirstName.LastName@aei.mpg.de}}

\vspace*{0.5cm}

Max-Planck-Institut f{\"u}r Gravitationsphysik\\
Albert-Einstein-Institut\\
Am M{\"u}hlenberg 1\\
14476 Golm, Germany\\
\vspace*{3cm}

{\bf Abstract}
\end{center}

We consider the limit of two-dimensional $N= (2,2)$ superconformal
minimal models when the central charge approaches $c=3$. Starting from
a geometric description as non-linear sigma models, we show that one
can obtain two different limit theories. One is the free theory of two
bosons and two fermions, the other one is a continuous orbifold
thereof. We substantiate this claim by detailed conformal field theory
computations. 

\newpage

\tableofcontents

\section{Introduction}

Sequences of two-dimensional conformal field theories and their limits
have been analysed
in~\cite{Runkel:2001ng,Graham:2001tg,Roggenkamp:2003qp,Fredenhagen:2004cj,Fredenhagen:2007tk,Roggenkamp:2008jm,Fredenhagen:2010zh,Fredenhagen:2012rb}. The
motivation to study them arises on the one hand, because one can use
them to explore non-rational models that occur as limits of sequences
of rational theories. On the other hand, the analysis of such
sequences might shed light on the structure of the space of
two-dimensional field theories. There are several ideas about what a good
notion of distance on such a space would be, e.g.\ the Zamolodchikov
metric~\cite{Zamolodchikov:1986gt,Kutasov:1988xb} or g-factors of
conformal interfaces (for a recent discussion
see~\cite{Douglas:2010ic}). If one has a notion of distance or more
generally a topology, one can also discuss the question of convergence
of sequences of theories. Conversely, lacking a proper understanding
of theory space, one may study sequences of theories in order to learn
more about what the right notion of convergence should be.

The study of limits of sequences of two-dimensional conformal field
theories was pioneered in \mbox{ref.\ \cite{Runkel:2001ng}} for the
limit of Virasoro minimal models when the central charge approaches
$c=1$. The general idea of that construction is to define fields in
the limit theory as limits of averages of fields. More precisely,
given a smooth, non-negative function~$f(h)$ of fast enough decay that
describes a certain averaging over conformal weights, the
corresponding limit field $\Phi_{\!f}$ arises from weighted averages
\begin{equation}\label{generalaverage}
\Phi_{\!f}^{(k)} = \sum_{i} f (h_{i}) \,\phi^{(k)}_{i}
\end{equation}
of primary fields $\phi^{(k)}_{i}$ in the $k^{\text{th}}$ model with
conformal weight $h_{i}$. Correlators of limit fields are then defined
as limits of correlators of averaged fields. In order to obtain finite
and sensible correlators in the limit, one can make use of the freedom
to rescale fields and correlators and to redefine the fields
$\phi^{(k)}_{i}$ by individual phases. A priori it is not guaranteed
that in this way one arrives at a valid conformal field theory, but in
all examples that have been studied over the years, the limit theory
seems to be well behaved. 

The above procedure is ambiguous when other quantum numbers are
available, because there is some freedom of how to treat them while
taking the limit. In the limit of $N=(2,2)$ superconformal minimal
models at central charge $c=3$, which we explore here, we will
encounter this ambiguity because of the presence of the $U(1)$ current
$J$ in the $N=2$ superconformal algebra and the corresponding charge
$Q$. On the one hand we could keep the charge fixed in the limit and
define field averages 
\begin{equation}\label{firstaverage}
^{(1)}\Phi^{(k)}_{\!f} = \sum_{i} f (h_{i},Q_{i})\, \phi^{(k)}_{i} \ ,
\end{equation}
corresponding to a certain test function $f(h,Q)$ by an obvious generalisation
of~\eqref{generalaverage}. This leads to the limit theory constructed
in~\cite{Fredenhagen:2012rb} with a continuous spectrum of charged
primary fields. On the other hand, one could rescale the charges and
define new averaged fields
\begin{equation}\label{scaledchargeaverage}
^{(2)}\Phi^{(k)}_{\!f} = \sum_{i} f(h_{i},Q_{i}(k+2)) \,\phi^{(k)}_{i} \ .
\end{equation}
Because of the rescaling, in this theory the primary fields have
charge zero, and we will show that this limit is equivalent to a free
theory of two uncompactified bosons and two fermions. The discrete
quantum number that arises from the rescaled charge of the primary
fields is then interpreted as the eigenvalue of the rotation operator
on the plane spanned by the two bosonic fields. In the process of
defining the limit theory we will see that in addition to a global
rescaling of the fields we also have to make use of the freedom to
redefine the ingredient fields $\phi^{(k)}_{i}$ of
$^{(2)}\Phi^{(k)}_{\!f}$ by individual phases compared to the
conventions used for the ingredient fields in the definition of
$^{(1)}\Phi^{(k)}_{\!f}$ in the other limit
construction~\cite{Fredenhagen:2012rb}.  
\smallskip

The appearance of two different limit theories can also be understood
from a geometric point of view. The minimal models can be described by
non-linear sigma models~\cite{Maldacena:2001ky} with a target space
having the topology of a disc with infinite circumference but with a
finite radius~$\sqrt{\pi(k+2)/2}$, which goes to infinity in the limit
$k\to\infty$. If one focusses on the region around the centre while
taking the limit, the metric approaches the flat metric on the
plane. This is the free limit theory described above. On the other
hand one could focus on the region close to the (singular) boundary of
the disc. As explained in section~\ref{sec:geometry} one can use
T-duality to show that the corresponding limit theory should be given
by a free theory of two bosons and two fermions orbifolded by the
rotation group $SO(2)$. We will verify explicitly that this continuous
orbifold coincides with the limit theory involving the fields
$^{(1)}\Phi^{(k)}_{\!f}$ that was constructed
in~\cite{Fredenhagen:2012rb}.  
\smallskip

The plan of the paper is the following: We will start our analysis in
section~\ref{sec:geometry} by discussing the two possible limits
starting from a geometric description. In section~\ref{sec:freefields}
we will confirm that the limit procedure using the fields
$^{(2)}\Phi^{(k)}_{\!f}$ (see~\eqref{scaledchargeaverage}) with
rescaled charges leads to a free field
theory; we determine the partition function, the three-point function
and boundary conditions in the limit and compare them to the free
field theory results. Thereafter in section~\ref{sec:contorbi} we show
that the other limit theory involving the fields
$^{(1)}\Phi^{(k)}_{\!f}$ (see~\eqref{firstaverage}) is
equivalent to a continuous orbifold by matching the partition function
and boundary conditions, and we close in section~\ref{sec:discussion}
with a brief discussion.

\section{The geometry of minimal models and the limit}
\label{sec:geometry}

In~\cite{Maldacena:2001ky} Maldacena, Moore and Seiberg gave a
geometric description of $N= (2,2)$ minimal models in terms of a
supersymmetric non-linear sigma model on a two-dimensional target
space with the topology of a disc and with the metric
\begin{equation}\label{mmgeometry}
ds^{2} = \frac{k+2}{1-\rho^{2}} \left(d\rho^{2} + \rho^{2}d\varphi^{2}
\right)\ ,
\end{equation}
where the radial coordinate $\rho$ runs from $0$ to $1$, and the
angular coordinate $\varphi$ is $2\pi$-periodic. In addition there is
a non-trivial dilaton field $\Phi$ of the form
\begin{equation}
e^{\Phi (\rho ,\varphi) -\Phi_{0}} = \frac{1}{\sqrt{1- \rho^{2}}} \ .
\end{equation}
The geometry is such that the boundary at $\rho =1$ is at a finite
distance $\frac{\pi}{2}\sqrt{k+2}$ from the centre at $\rho =0$, but the
circumference of a circle at radius $\rho$ is $2\pi
\rho\sqrt{\frac{k+2}{1-\rho^{2}}}$, and it diverges as $\rho \to 1$.

Given the geometric interpretation, we now want to analyse what
happens for large levels $k$. One way to take the geometric limit is
to introduce a new coordinate 
\begin{equation}
\rho '=\sqrt{k+2}\,\rho \ ,
\end{equation}
such that the
metric reads
\begin{equation}
ds^{2} = \frac{1}{1-\rho '^{2}/ (k+2)}\left(d\rho'^{2}+\rho '^{2}d\varphi
^{2}\right) \ .
\end{equation}
Keeping $\rho '$ fixed while taking the limit $k\to \infty$ leads to
the flat metric on the plane.

From this analysis one would like to conclude that the limit of $N=
(2,2)$ minimal models for $k\to\infty$ is a free theory. At first
sight this is in conflict with the analysis in~\cite{Fredenhagen:2012rb}, where
the limit of minimal models was shown to be a theory containing fields
with a continuous $U (1)$ charge that should not be present in a free
theory. This conflict can be resolved by comparing more carefully how
the limits are taken in these two approaches.

In a minimal model of level $k$, Neveu-Schwarz primary fields
$\phi_{l,m}$ are labelled by two integers satisfying $0\leq l\leq k$,
$|m|\leq l$ and $l+m$ even. In~\cite{Fredenhagen:2012rb}
the fields in the limit theory arise from fields $\phi_{l,m}$ where
$l$ and $m$ grow linearly with $k$ in the limit, while the
difference $l-|m|=:2n$ is kept fixed. To compare this procedure to the
geometric limit we need a geometric interpretation of the fields
$\phi_{l,m}$, which was also given in~\cite{Maldacena:2001ky}. The
fields $\phi_{l,m}$ correspond to wavefunctions
\begin{equation}
\psi_{l,m} (\rho ,\varphi) = \rho^{|m|} e^{im\varphi} {}_{2}F_{1}
\left(\tfrac{|m|+l}{2}+1,\tfrac{|m|-l}{2};|m|+1;\rho^{2}\right)\ ,
\end{equation}
which are eigenfunctions of the (dilaton-corrected) Laplacian,
\begin{equation}
\left( -\frac{1}{2}\nabla^{2} + (\nabla \Phi)\cdot \nabla\right)\psi_{l,m}
(\rho ,\varphi) = 2h_{l,m} \,\psi_{l,m} (\rho ,\varphi) \ .
\end{equation}
Here, ${}_{2}F_{1}$ is the hypergeometric function, and  
\begin{equation}
h_{l,m} = \frac{l (l+2)-m^{2}}{4 (k+2)} 
\end{equation}
is the conformal weight of the field $\phi_{l,m}$.

When we now take the geometric limit to the flat plane, the
wavefunctions $\psi_{l,m}$ should approach the eigenfunctions of the flat
Laplacian. In radial coordinates, these are given by 
\begin{equation}
\psi^{\text{flat}}_{p,m} (\rho ',\varphi) = e^{im\varphi} J_{|m|} (p\rho ')\ , 
\end{equation}
where $J$ is a Bessel function of the first kind. They satisfy
\begin{equation}
-\frac{1}{2}\,\nabla_{\!\text{flat}}^{2} \psi^{\text{flat}}_{p,m} (\rho ',\varphi) =
\frac{p^{2}}{2}\,\psi^{\text{flat}}_{p,m} (\rho ',\varphi) \ .
\end{equation}
Comparing the angular dependence of $\psi_{l,m}$ and $\psi^{\text{flat}}_{p,m}$
one observes immediately that the label $m$ should be kept fixed in
the limit. For the eigenvalue $h_{l,m}$ to approach
$h_{p}=\frac{p^{2}}{4}$, the label $l$ has to grow with the square
root of $k$, namely $l\approx p \sqrt{k+2}$. Then the wavefunctions
$\psi_{l,m}$ behave as
\begin{align}
(k+2)^{|m|/2}\psi_{l,m} &= \rho '^{|m|}e^{im\varphi}\, {}_{2}F_{1}
(\tfrac{|m|+l}{2}+1,\tfrac{|m|-l}{2};|m|+1;\tfrac{\rho'^{2}}{k+2})\\
& = e^{im\varphi} \sum_{n=0}^{(l -|m|)/2} \frac{\left(\frac{l+|m|}{2}
\right)_{n} \left(\frac{-l+|m|}{2} \right)_{n}}{n!(|m|+1)_{n}}
(\rho')^{2n+|m|} (k+2)^{-n} \\
& = e^{im\varphi} \sum_{n=0}^{(l-|m|)/2} \frac{(-1)^{n}}{n!
(|m|+1)_{n}} \left(\frac{l-|m|-2n+2}{2\sqrt{k+2}} \right) \dotsb
\left(\frac{l-|m|}{2\sqrt{k+2}} \right)\nonumber\\
& \qquad \times 
\left(\frac{l+|m|}{2\sqrt{k+2}} \right) \dotsb
\left(\frac{l+|m|+2n-2}{2\sqrt{k+2}} \right) (\rho')^{2n+|m|} \\
& \sim  e^{im\varphi} \sum_{n=0}^{\infty} 
\frac{(-1)^{n}p^{2n}2^{-2n}}{n!(|m|+1)_{n}} (\rho')^{2n+|m|}\\
& \sim  e^{im\varphi} \,J_{|m|} (p\rho') \ .
\end{align}
Thus up to an overall normalisation factor the wavefunctions
$\psi_{l,m}$ approach the wavefunctions of the free theory.

On the one hand this suggests that there is a free field theory limit
of minimal models by scaling $l\approx p\sqrt{k+2}$ and keeping $m$
fixed. This will be examined further in section~\ref{sec:freefields}. On the
other hand this means that the limit theory found in~\cite{Fredenhagen:2012rb}
should correspond to a different way of taking the geometric limit.
Indeed for fixed $l-|m|=2n$ the wavefunction $\psi_{l,m}$ is, apart from
the angular part, a polynomial in $\rho$ containing $n+1$ terms with
powers ranging from $\rho^{|m|}$ to $\rho^{|m|+2n}$. If $|m|$ is
large, the wavefunctions are localised close to $\rho =1$, in the
region where the metric and the dilaton diverge and the sigma model
description becomes singular, so that one cannot easily extract a
sensible geometric interpretation. It was however observed
in~\cite{Maldacena:2001ky} that under a T-duality the minimal
model is mapped to its own $\mathbb{Z}_{k+2}$ orbifold described by
\begin{align}
d\tilde{s}^{2} & =
\frac{k+2}{1-\tilde{\rho}^{2}}\left(d\tilde{\rho}^{2}+\tilde{\rho}^{2}d\tilde{\varphi}^{2}\right)\\
e^{\tilde{\Phi} -\Phi_{0}} & = \frac{1}{\sqrt{k+2}}
\frac{1}{\sqrt{1-\tilde{\rho}^{2}}} \\
\tilde{\varphi} & \equiv  \tilde{\varphi} +\frac{2\pi}{k+2} \ .
\label{orbifoldidentification}
\end{align}
T-duality maps the problematic region around $\rho =1$ to the
region close to the conical singularity of the orbifold at
$\tilde{\rho}=0$. This suggests that the limit of minimal models
of~\cite{Fredenhagen:2012rb} corresponds to taking the limit in the orbifolded
model by focussing on the region around $\tilde{\rho}=0$. By
introducing again a rescaled variable
$\tilde{\rho}'=\sqrt{k+2}\tilde{\rho}$ and keeping $\tilde{\rho}'$
fixed in the limit, the metric $d\tilde{s}^{2}$ approaches the flat
metric on the plane. On the other hand, according
to~\eqref{orbifoldidentification} all angles have to be
identified. The resulting limit theory is thus the theory on a flat
plane $\mathbb{R}^{2}$ orbifolded by the rotation group $SO (2)$.

In section~\ref{sec:contorbi} we will construct this orbifold conformal
field theory and show that it precisely matches the limit theory
of~\cite{Fredenhagen:2012rb}.

\section{Free field limit}
\label{sec:freefields}

The geometric analysis of section~\ref{sec:geometry} suggests that the
$N=2$ minimal models have a free field limit when the labels of the
Neveu-Schwarz primary fields $\phi_{l,m}$ are treated such that
$l\approx \sqrt{k+2}\,p$ and $m$ stays fixed in the limit. As
$m=-(k+2)Q$ for these fields, we are led to consider the limit of
averaged fields $^{(2)}\Phi^{(k)}_{\!f}$ that uses a charge rescaled
by $(k+2)$ (see~\eqref{scaledchargeaverage}). We will first analyse
the behaviour of the partition function in the limit. We will then
turn to the actual construction of the fields in the limit theory, and
determine the bulk three-point function and boundary conditions.

\subsection{Partition function}

We will now reproduce the partition function of the free theory of two
uncompactified bosons and two fermions as the limit of the partition
functions of minimal models. We focus on the Neveu-Schwarz sector, and
for the minimal models we define
\begin{equation}
\mathcal{P}^{\text{NS}}_{k} (\tau,\nu) = \Tr_{\cH_{k}^{\text{NS}}}
\left( q^{L_{0}-\frac{c}{24}}z^{J_{0}}\,
\bar{q}^{\bar{L}_{0}-\frac{c}{24}}\bar{z}^{\bar{J}_{0}} \right)\ ,
\end{equation}
where $J_{0}$ is the zero mode of the $U(1)$ current of the $N=2$
superconformal algebra, and $q=e^{2\pi i\tau}$ and $z=e^{2\pi
i\nu}$. Note that $\mathcal{P}^{\text{NS}}_{k} (\tau,\nu)$ does not depend
holomorphically on $\tau$ and $\nu$, but we suppress the dependence on
$\bar{\tau}$ and $\bar{\nu}$ to shorten the notation. $\cH^{\text{NS}}_{k}$ is the
full supersymmetric Hilbert space for the Neveu-Schwarz sector,
\begin{equation}
\mathcal{H}^{\text{NS}}_{k} = \bigoplus_{0\leq l\leq k}\bigoplus_{\substack{|m|\leq l\\ l+m\ \text{even}}}
\mathcal{H}_{l,m}\otimes \mathcal{H}_{l,m} \ ,
\end{equation}
and the Neveu-Schwarz spectrum of the actual minimal model corresponds to a (GSO-like)
projection thereof. For $k\to\infty$ the partition function diverges:
there are infinitely many states approaching the same conformal weight
and charge. This can be seen by looking at the contribution of the
Neveu-Schwarz ground states,
\begin{equation}
\mathcal{P}^{\text{NS,g.s.}}_{k} (\tau,\nu) =  
\sum_{\substack{|m|\leq l \leq k \\ l+m\ \text{even}}}
(q\bar{q})^{\frac{(l+1)^{2}-m^{2}}{4 (k+2)}}
(z\bar{z})^{-\frac{m}{k+2}}\ , 
\end{equation}
where we sum over the leading term of the minimal model characters
given in \mbox{eq.\ \eqref{minmod-NS-character}}. By introducing the
summation variable $n=\frac{1}{2} (l-|m|)$, we can rewrite the sum and
perform the summation over $m$,
\begin{align}
\mathcal{P}^{\text{NS,g.s.}}_{k} (\tau,\nu) & = \sum_{n=0}^{\lfloor
\frac{k}{2}\rfloor}  (q\bar{q})^{\frac{(2n+1)^{2}}{4 (k+2)}}
\left(\sum_{m=0}^{k-2n} + \sum_{m=-k+2n}^{-1} \right) 
(q\bar{q})^{\frac{(2n+1)|m|}{2(k+2)}}(z\bar{z})^{-\frac{m}{k+2}}\\
&=  \sum_{n=0}^{\lfloor
\frac{k}{2}\rfloor}(q\bar{q})^{\frac{(2n+1)^{2}}{4 (k+2)}}
\bigg[\frac{1-(q\bar{q})^{\frac{2n+1}{2 (k+2)}
(k-2n+1)}(z\bar{z})^{-\frac{k-2n+1}{k+2}}}{1
-(q\bar{q})^{\frac{2n+1}{2 (k+2)}}(z\bar{z})^{-\frac{1}{k+2}}}\nonumber\\
&\qquad \qquad \qquad \qquad  
+\frac{1-(q\bar{q})^{\frac{2n+1}{2 (k+2)}
(k-2n+1)}(z\bar{z})^{\frac{k-2n+1}{k+2}}}{1
-(q\bar{q})^{\frac{2n+1}{2 (k+2)}}(z\bar{z})^{\frac{1}{k+2}}} 
\bigg] \ .
\label{partfuncdiverges}
\end{align}
Here, $\lfloor x \rfloor$ denotes the greatest integer smaller or
equal $x$. Now let us cut off the summation over $n$ by $n\leq \Lambda
\sqrt{k+2}$ with $0<\Lambda <1$. In this summation range, the
denominators in~\eqref{partfuncdiverges} go to zero, and the summands
diverge as $\frac{k+2}{n+\dotsb}$. The sum over $n$ produces a
logarithmic divergence, such that the leading divergence of the
partition function is of the form $(k+2)\log(k+2)$. The divergence
signals an infinite degeneracy of states. Part of the divergence might be
resolved by introducing additional quantum numbers that lift the
degeneracy; on the other hand there can be a divergence due to the
emergence of a continuous spectrum, in which case we can regularise
the partition function by rescaling the density of states
appropriately.\footnote{Consider e.g.\ a free compact bosonic field
$\phi\equiv \phi +2\pi R$. For $R\to\infty$ the partition function
diverges as the volume $R$, and for the noncompact boson one usually considers
the regularised partition function rescaled by $1/R$.}

The fields we are interested in have fixed label $m$, and their $U(1)$
charges $Q=-\frac{m}{k+2}$ (the eigenvalues of $J_{0}$) approach zero
in the limit. To cure the divergence associated to the appearance of
infinitely many chargeless fields, we want to keep track of the
quantum number $m$ in the limit. In the free field theory, $m$
corresponds to the eigenvalue of the angular momentum operator $M$,
and we could insert $e^{i\varphi M}$ in the partition function: in
this way the partition function is written as a formal power series in
$e^{i\varphi}$ and $e^{-i\varphi}$, and the coefficient of
$e^{im\varphi}$ gives the contribution of states of a given angular
momentum $m$. In the geometric description of the minimal models,
there is a $U(1)$ rotation symmetry in the classical theory, but it is
broken to a $\mathbb{Z}_{k+2}$ symmetry in the quantum model. The
rotation by an angle $2\pi i\frac{r}{k+2}$ ($r$ integer) is realised
by the operator $g^{r}$, where $g$ acts on states in $\cH_{l,m}\otimes
\cH_{l,m}$ by multiplication with the phase $e^{2\pi
i\frac{m}{k+2}}$. To mimic the insertion of $e^{i\varphi M}$ in the
free field theory, we therefore introduce the operator $g^{\lfloor
\frac{\varphi}{2\pi} (k+2)\rfloor}$ in the partition function, such
that states with a given $m$ will get the phase $e^{im\varphi}$ in the
limit.

The regularised partition function therefore becomes (we use standard
conventions for $\vartheta$-functions as summarised in
appendix~\ref{app:characters})
\begin{equation}
\cP^{\text{NS}}_{k,(\varphi)} (\tau,\nu) = \left|\frac{\vartheta_3(\tau
,\nu)}{\eta^3(\tau)}\right|^2 \sum_{m=-k}^{k} e^{2\pi
i\frac{m}{k+2}\lfloor \frac{\varphi}{2\pi} (k+2)\rfloor} 
(z\bar{z})^{-\frac{m}{k+2}} \sum_{\substack{l=|m|\\ l+m\ \text{even}}}^{k}
(q\bar{q})^{\frac{(l+1)^{2}-m^{2}}{4 (k+2)}} 
\big|\Gamma_{lm}^{(k)} (\tau,\nu )\big|^{2} \ ,
\end{equation}
where we used the minimal model characters given in
\mbox{eq.\ \eqref{minmod-NS-character}}. $\Gamma_{lm}^{(k)}$ is defined
in \mbox{eq.\ \eqref{def-Gamma}}, it is of the form
\begin{equation}
\Gamma_{lm}^{(k)}=1+ (\text{subtractions from singular vectors}) \ ,
\end{equation}
and its behaviour for large $k$ is given in \mbox{eq.\ \eqref{limit-Gamma}}. 
The contribution of a fixed $m$ is then
\begin{equation}
\cP^{\text{NS}}_{k,(\varphi,m)} (\tau,\nu) \approx \left|\frac{\vartheta_3(\tau
,\nu)}{\eta^3(\tau)}\right|^2 e^{im\varphi} 
\frac{\sqrt{k+2}}{2} \int dp\, (q\bar{q})^{p^{2}/4} \ ,
\end{equation}
where we employed the Euler-MacLaurin sum formula (see
e.g.~\cite[appendix D]{Andrews:book}) to convert the sum over $l$ into
an integral over $p=l/\sqrt{k+2}$. For fixed $m$ and large $l$ all
singular vectors disappear and $\Gamma_{lm}^{(k)}\to 1$. 
To get the true partition function, i.e.\ the trace over the projected
Hilbert space, we have to combine $\cP$ evaluated at $\nu$ and at 
$\nu+i\pi$, and we find after rescaling by an overall factor
\begin{multline}
\frac{1}{\sqrt{k+2}} \left(\cP^{\text{NS}}_{k,(\varphi)} (\tau ,\nu) +
\cP^{\text{NS}}_{k,(\varphi)} (\tau ,\nu +i\pi) \right) \\
\to 
\frac{1}{2}\left(\left|\frac{\vartheta_3(\tau,\nu)}{\eta^3(\tau)}\right|^2
+\left|\frac{\vartheta_4(\tau,\nu)}{\eta^3(\tau)}\right|^2
\right) \sum_{m\in\mathbb{Z}} e^{im\varphi} \int_{0}^{\infty}dp\,
(q\bar{q})^{p^{2}/4} \ ,
\end{multline}
which is precisely the Neveu-Schwarz part of the partition function of
two free uncompactified bosons and two fermions~(see e.g.\
\cite[chapter 12.2]{BlumenhagenLuestTheisen}), weighted by the
rotation operator $e^{iM\varphi}$. The rescaling can be
explained from the analysis in the following subsection: in the
interval $[p,p+\Delta p]$ there are $\frac{\sqrt{k+2}}{2}\Delta p$
ground states contributing to the partition function
(see~\eqref{densofstates}). The rescaling therefore corresponds to
adjusting the density of states to $1$ per unit interval $\Delta p$. 

\subsection{Fields and correlators}

We will now define fields $\Phi_{p,m}$ in the limit theory, which
arise from averaged fields $^{(2)}\Phi^{(k)}_{\!f}$ with specific
averaging functions $f$. For $m=0$, the
behaviour of the corresponding fields in the limit was analysed
in~\cite{Fredenhagen:2012rb}, and we will closely follow that
construction. 

In the Neveu-Schwarz sector of the $k^{\text{th}}$ minimal model we introduce the averaged
fields\footnote{In comparison to the discussion
around~\eqref{scaledchargeaverage} we make use of the fact that the
spectrum of the rescaled charge $Q(k+2)=-m$ is discrete so that we can
define fields with fixed labels $m$. For large $k$ our procedure here then
corresponds to using a (discontinuous) averaging function\\
$f_{p,\epsilon} (h)=\left\{\!\begin{array}{ll}
\!1/\epsilon & \text{for}\ |p-2\sqrt{h}|<\epsilon /2\\
\!0 & \text{else}
\end{array} \right.$, and in addition a $k$-dependent rescaling of the
fields by $2/\sqrt{k+2}$.}
\begin{equation}
\Phi^{\epsilon ,k}_{p,m} = \frac{1}{\left|N (p,\epsilon
,k,m)\right|} 
\sum_{l\in N (p,\epsilon ,k,m)} \phi_{l,m} \ ,
\end{equation}
where $\phi_{l,m}$ are Neveu-Schwarz primary fields (labelled by two
integers with $0\leq l\leq k$, $|m|\leq l$ and $l+m$ even), and the
set $N (p,\epsilon ,k,m)$ contains all allowed labels $l$ that are
close to $p\sqrt{k+2}$,
\begin{equation}
N (p,\epsilon ,k,m) = \left\{ l: l+m\ \text{even}\, ,\ p-\frac{\epsilon}{2} <
\frac{l}{\sqrt{k+2}} < p + \frac{\epsilon}{2}  \right\} \ .
\end{equation}
Here, $\epsilon$ is a small real number that will be taken to zero at
the end. For large $k$ the number of elements in  $N (p,\epsilon ,k,m)$
is (assuming $p-\frac{\epsilon}{2}>0$)
\begin{equation}\label{densofstates}
\left|N (p,\epsilon ,k,m)\right| = \epsilon \frac{\sqrt{k+2}}{2} + \mathcal{O} (1) \ .
\end{equation}
These averaged fields are used to define fields $\Phi_{p,m}$ in the
limit theory of conformal weight $h=\frac{p^{2}}{4}$ and $U (1)$
charge $Q=0$. Their correlators are defined as
\begin{multline}
\langle \Phi_{p_{1},m_{1}} (z_{1},\bar{z}_{1}) \dotsb \Phi_{p_{r},m_{r}} (z_{r},\bar{z}_{r})\rangle\\
= \lim_{\epsilon \to 0} \lim_{k\to \infty} 
\beta (k)^{2} \alpha (k)^{r}\langle \Phi_{p_{1},m_{1}}^{\epsilon ,k}
(z_{1},\bar{z}_{1})\dotsb \Phi_{p_{r},m_{r}}^{\epsilon ,k} (z_{r},\bar{z}_{r})\rangle \ ,
\end{multline}
with normalisation factors $\alpha (k)$ for each field, and an overall 
normalisation factor $\beta^{2} (k)$ for correlators on the sphere. In
addition to this rescaling we also have the possibility to redefine
the fields $\phi_{l,m}$ by individual phases. Compared to the analysis
in~\cite{Fredenhagen:2012rb} we change the normalisation by 
\begin{equation}\label{conventionchange}
\phi_{l,m} \to (-1)^{\frac{l-m}{2}} \phi_{l,m} \ .
\end{equation}
The necessity of introducing these signs will become clear when we
analyse the three-point function. With this convention
the two-point function in the minimal models is
\begin{align}
\langle \phi_{l_{1},m_{1}} (z_{1},\bar{z}_{1})\phi_{l_{2},m_{2}}
(z_{2},\bar{z}_{2})\rangle &=
(-1)^{\frac{l_{1}-m_{1}+l_{2}-m_{2}}{2}}\,
\delta_{l_{1},l_{2}}\, \delta_{m_{1}+m_{2},0} \,\frac{1}{|z_{12}|^{4h_{1}}} \nonumber\\
&= (-1)^{m_{1}}\, \delta_{l_{1},l_{2}}\, \delta_{m_{1}+m_{2},0}\,
\frac{1}{|z_{12}|^{4h_{1}}} \ ,
\end{align}
where we used that $l_{1}+m_{1}$ is even.

By following the analysis of~\cite{Fredenhagen:2012rb} we find a normalised
two-point function in the limit,
\begin{equation}
\left\langle \Phi_{p_{1},m_{1}}
(z_{1},\bar{z}_{1})\Phi_{p_{2},m_{2}} (z_{2},\bar{z}_{2})\right\rangle
= (-1)^{m_{1}}\delta (p_{1}-p_{2}) \delta_{m_{1}+m_{2},0}\frac{1}{|z_{1}-z_{2}|^{p_{1}^{2}}}  \ ,
\end{equation}
if we choose
\begin{equation}\label{alphabeta}
\alpha(k)\beta (k) = \frac{(k+2)^{1/4}}{\sqrt{2}} \ .
\end{equation}
Before moving on, let us compare this to the free field theory of two
uncompactified bosons and two fermions. The
primary fields $\Phi^{\text{free}}_{\mathbf{p}}$ in the Neveu-Schwarz
sector are labelled by a complex momentum $\mathbf{p}$, they have
conformal weight $h=\frac{|\mathbf{p}|^{2}}{4}$ and $U (1)$ charge
$q=0$. We can define a new ``radial'' basis,
\begin{equation}
\Phi^{\text{free}}_{p,m} = \sqrt{\frac{p}{2\pi}}\int d\varphi\ 
\Phi^{\text{free}}_{pe^{i\varphi}}\, e^{im\varphi} \ ,
\end{equation}
where the factor in front ensures a proper normalisation of the
two-point function,
\begin{equation}
\langle \Phi^{\text{free}}_{p_{1},m_{1}}
(z_{1},\bar{z}_{1})\Phi^{\text{free}}_{p_{2},m_{2}} (z_{2},\bar{z}_{2})\rangle
= (-1)^{m_{1}}\delta (p_{1}-p_{2}) \delta_{m_{1}+m_{2},0}
\frac{1}{|z_{1}-z_{2}|^{p_{1}^{2}}} \ .
\end{equation}
We therefore expect that the fields $\Phi_{p,m}$ of the limit theory
are to be identified with the fields $\Phi^{\text{free}}_{p,m}$ of the
free field theory. To confirm this we now look at the three-point
function.

\subsection{Three-point function}

The three-point function in the free theory is given by
\begin{multline}
\langle \Phi^{\text{free}}_{\mathbf{p}_{1}} (z_{1},\bar{z}_{1})
\Phi^{\text{free}}_{\mathbf{p}_{2}} (z_{2},\bar{z}_{2})
\Phi^{\text{free}}_{\mathbf{p}_{3}} (z_{3},\bar{z}_{3})\rangle = 
\delta^{(2)} (\mathbf{p}_{1}+\mathbf{p}_{2}+\mathbf{p}_{3})\\
\times  
|z_{12}|^{2 (h_{3}-h_{1}-h_{2})} |z_{23}|^{2 (h_{1}-h_{2}-h_{3})} 
 |z_{13}|^{2 (h_{2}-h_{1}-h_{3})} \ . 
\label{freefield}
\end{multline}
A straightforward calculation (see appendix~\ref{sec:app-free-three}) shows that
in the basis $\Phi^{\text{free}}_{p,m}$ it can be expressed as
\begin{multline}
\langle \Phi^{\text{free}}_{p_{1},m_{1}} (z_{1},\bar{z}_{1})
\Phi^{\text{free}}_{p_{2},m_{2}} (z_{2},\bar{z}_{2})
\Phi^{\text{free}}_{p_{3},m_{3}} (z_{3},\bar{z}_{3})\rangle
= \delta_{m_{1}+m_{2}+m_{3},0}
\frac{\sqrt{p_{1}p_{2}p_{3}}}{\sqrt{2\pi}}
(-1)^{m_{3}}\\
\times 
\frac{\cos (m_{2}\alpha_{1}-m_{1}\alpha_{2})}{A (p_{1},p_{2},p_{3})}
|z_{12}|^{2 (h_{3}-h_{1}-h_{2})} |z_{23}|^{2 (h_{1}-h_{2}-h_{3})} 
 |z_{13}|^{2 (h_{2}-h_{1}-h_{3})} \ ,
\label{freefieldthreepointfn}
\end{multline}
where $A (p_{1},p_{2},p_{3})$ is the area of the triangle with side lengths
$p_{1}$, $p_{2}$ and $p_{3}$, and $\alpha_{i}$ is the angle of the
triangle opposite of the edge $p_{i}$. If a triangle with these side
lengths does not exist, the correlator is zero.

The three-point functions in the limit theory are obtained from the
three-point functions in the minimal models~\cite{Mussardo:1988av}
(see also~\cite{Zamolodchikov:1986bd,Dotsenko:1990zb}). For large $k+2$
and $l_{i}\approx p_{i}\sqrt{k+2}$, the three-point function is given
by (see~\cite{Fredenhagen:2012rb})
\begin{multline}
\langle\phi_{l_{1},m_{1}} (z_{1},\bar{z}_{1}) \phi_{l_{2},m_{2}}
(z_{2},\bar{z}_{2})\phi_{l_{3},m_{3}} (z_{3},\bar{z}_{3})\rangle
= (-1)^{\frac{l_{1}+l_{2}+l_{3}}{2}} \begin{pmatrix}
\frac{l_{1}}{2} & \frac{l_{2}}{2} & \frac{l_{3}}{2}\\
\frac{m_{1}}{2} & \frac{m_{2}}{2}& \frac{m_{3}}{2}
\end{pmatrix}^{2}\\
\times \sqrt{(l_{1}+1)(l_{2}+1)(l_{3}+1)} \, 
\delta_{m_{1}+m_{2}+m_{3},0}|z_{12}|^{2(h_{3}-h_{1}-h_{2})}
|z_{13}|^{2(h_{2}-h_{1}-h_{3})}|z_{23}|^{2(h_{1}-h_{2}-h_{3})}
\ .
\end{multline}
Here, $\begin{pmatrix}
j_{1}&j_{2}&j_{3} \\
\mu_{1}&\mu_{2}&\mu_{3}
\end{pmatrix}$ denotes the Wigner 3j-symbols, and in order to determine
the correlator in the limit one has to understand the asymptotic behaviour of
the 3j-symbols for large quantum numbers $j_{i}$, which we analyse in
appendix~\ref{app:3j}. The result (compare with~\eqref{3j-asympt-app}) is
\begin{multline}
\begin{pmatrix}
\frac{l_{1}}{2} & \frac{l_{2}}{2} & \frac{l_{3}}{2} \\
\frac{m_{1}}{2} & \frac{m_{2}}{2} & \frac{m_{3}}{2}
\end{pmatrix} 
= (k+2)^{-1/2}\frac{(-1)^{\frac{l_{1}-l_{2}-m_{3}}{2}}}{\sqrt{\frac{\pi}{2}A
(p_{1},p_{2},p_{3})}}\\
\times   \cos \left(\frac{l_{1}+l_{2}-l_{3}}{4}\pi
+\frac{m_{2}\alpha_{1}-m_{1}\alpha_{2}}{2} \right) + \mathcal{O} (k^{-1}) \ .
\end{multline}
For large level $k$ the three-point function therefore behaves as 
\begin{multline}
\langle\phi_{l_{1},m_{1}} (z_{1},\bar{z}_{1}) \phi_{l_{2},m_{2}}
(z_{2},\bar{z}_{2})\phi_{l_{3},m_{3}} (z_{3},\bar{z}_{3})\rangle
= (k+2)^{-1/4} \frac{2\sqrt{p_{1}p_{2}p_{3}}}{\pi A
(p_{1},p_{2},p_{3})} \delta_{m_{1}+m_{2}+m_{3},0}\\
\times (-1)^{\frac{l_{1}+l_{2}+l_{3}}{2}}
\cos^{2} \left(\tfrac{l_{1}+l_{2}-l_{3}}{4}\pi
+\tfrac{m_{2}\alpha_{1}-m_{1}\alpha_{2}}{2} \right) 
|z_{12}|^{2(h_{3}-h_{1}-h_{2})}
|z_{13}|^{2(h_{2}-h_{1}-h_{3})}|z_{23}|^{2(h_{1}-h_{2}-h_{3})}
\ .
\end{multline}
To obtain the correlator in the limit theory we have to average over
the quantum numbers~$l_{i}$. We observe that 
\begin{multline}
(-1)^{\frac{l_{1}+l_{2}+l_{3}}{2}}
\cos^{2} \left(\tfrac{l_{1}+l_{2}-l_{3}}{4}\pi
+\tfrac{m_{2}\alpha_{1}-m_{1}\alpha_{2}}{2} \right) \\
= (-1)^{m_{3}}
\times \left\{\begin{array}{ll}
\cos^{2} \left( \frac{m_{2}\alpha_{1}-m_{1}\alpha_{2}}{2} \right) &
\ \text{for}\ \frac{l_{1}+l_{2}-l_{3}}{2} = 0 \ \text{mod}\  2\\[5pt]
-\sin^{2} \left( \frac{m_{2}\alpha_{1}-m_{1}\alpha_{2}}{2} \right) &
\ \text{for}\ \frac{l_{1}+l_{2}-l_{3}}{2} = 1 \ \text{mod}\  2 \ . 
\end{array} \right. 
\end{multline}
In average these contributions combine to
\begin{equation}\label{average}
\frac{1}{2} (-1)^{m_{3}} \left(\cos^2
\tfrac{m_{2}\alpha_{1}-m_{1}\alpha_{2}}{2} -
\sin^2 \tfrac{m_{2}\alpha_{1}-m_{1}\alpha_{2}}{2}\right) = 
\frac{1}{2} (-1)^{m_{3}} \cos \left(
m_{2}\alpha_{1}-m_{1}\alpha_{2}\right) .
\end{equation} 
In total we arrive at
\begin{multline}
\langle\Phi_{p_{1},m_{1}} (z_{1},\bar{z}_{1}) \Phi_{p_{2},m_{2}}
(z_{2},\bar{z}_{2})\Phi_{p_{3},m_{3}} (z_{3},\bar{z}_{3})\rangle
 = \beta^{2}(k)\alpha^{3}(k) (k+2)^{-1/4} \,\delta_{m_{1}+m_{2}+m_{3},0} \,(-1)^{m_{3}} \\
\times \frac{\sqrt{p_{1}p_{2}p_{3}}}{\pi}\,\frac{\cos (m_{2}\alpha_{1}-m_{1}\alpha_{2})}{A
(p_{1},p_{2},p_{3})}\,
|z_{12}|^{2 (h_{3}-h_{1}-h_{2})} |z_{23}|^{2 (h_{1}-h_{2}-h_{3})} 
 |z_{13}|^{2 (h_{2}-h_{1}-h_{3})} \ ,
\end{multline}
which matches the free field theory
result~\eqref{freefieldthreepointfn} if we set
(respecting~\eqref{alphabeta})
\begin{align}
\alpha (k) &= \sqrt{2\pi} (k+2)^{-1/4} &
\beta (k) &= \frac{1}{2\sqrt{\pi}} (k+2)^{1/2} \ .
\end{align}
Hence, we find perfect agreement for the three-point function. Notice
that the redefinition of the minimal model fields $\phi_{l,m}$ by the
sign $(-1)^{\frac{l-m}{2}}$ was crucial in matching the
expressions. Without it, the averaging in~\eqref{average} would simply
give $\frac{1}{2} (-1)^{m_{3}}$ so that the three-point function would
have a rather trivial dependence on the labels $m_{i}$.

\subsection{A-type boundary conditions}

\begin{figure}
\begin{center}
\includegraphics{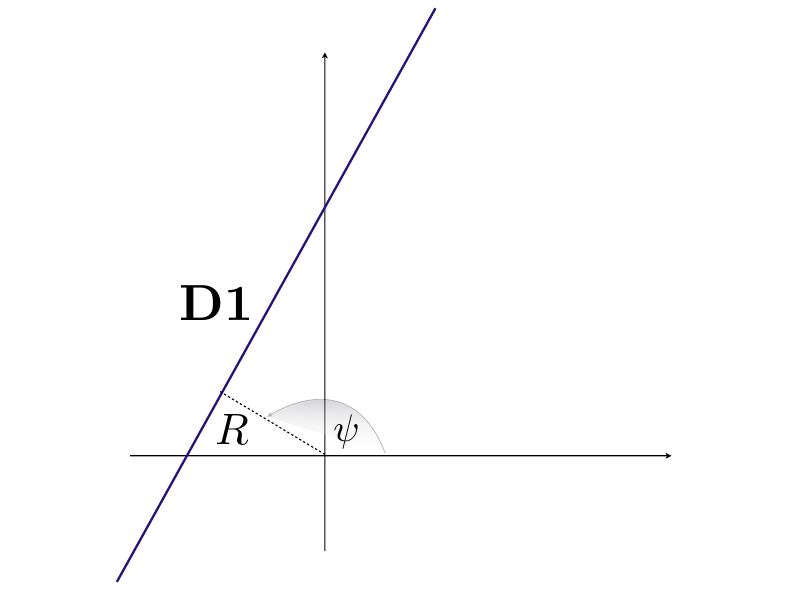}
\end{center}
\caption{\label{fig:coord-free}Illustration of the boundary condition
that corresponds to a one-dimensional brane, and the distance $R$ and
the angle $\psi$ that determine its position.}
\end{figure}

We now want to discuss boundary conditions. In a free theory the
simplest boundary conditions we can discuss are combinations of Dirichlet and Neumann
boundary conditions, and interpreting the free bosonic fields as
coordinates of a flat target space, such boundary conditions can be
formulated by specifying a flat submanifold (brane) that encodes the
possible boundary values of the fields. 
We first focus on one-dimensional branes in our two-dimensional
target. By choosing boundary conditions also for the
fermions, we can ensure appropriate boundary conditions for the
supercurrents such that the maximal amount of supersymmetry is
preserved. For our one-dimensional branes, this leads to
A-type gluing conditions for the supercurrents (for a discussion of
A- and B-type gluing conditions in $N=2$ supersymmetric field
theories see e.g.\ \cite{Ooguri:1996ck,Hori:2000ck}). In the free
theory, a one-dimensional brane is characterised by a vector
$Re^{i\psi}$ that determines its shortest distance from the origin
plus an orientation (see fig.\ \ref{fig:coord-free}). In the
Neveu-Schwarz sector the one-point functions are\footnote{\label{fn:branes}A boundary
condition corresponding to a $d$-dimensional brane in a
$D$-dimensional target space that only couples to the NS-NS sector has
the one-point function
\[
\big\langle e^{i\vec{p}\cdot\vec{X}}\big\rangle = 2^{-\frac{D}{4}} (\alpha
')^{\frac{D-2d}{4}} \delta^{(d)} (\vec{p}_{\parallel})e^{i\vec{R}\cdot
\vec{p}_{\perp}} |z-\bar{z}|^{-2h_{p}}\ ,
\]
where the conformal weight is $h_{p}=\frac{\alpha'p^{2}}{4}$. The
projection of the full Hilbert space only allows either the even- or
the odd-dimensional branes to couple to the R-R sector; in that case
there is an additional factor of $2^{-1/2}$. In our conventions $\alpha'=1$.
}
\begin{equation}
\big\langle \Phi^{\text{free}}_{pe^{i\varphi}}
(z,\bar{z})\big\rangle^{\!A}_{\!R,\psi} = \frac{1}{2}\,
\delta (p \cos (\psi -\varphi)) \, e^{iR p \sin (\psi -\varphi)}
\frac{1}{|z-\bar{z}|^{2h_{p}}} \ .
\end{equation}
The prefactor $1/2$ already includes the factor of $2^{-1/2}$ that
arises because we choose the (GSO-like)
projection of our theory such that also the Ramond-Ramond fields
couple to the one-dimensional brane. In the radial basis, the
one-point function is then given by
\begin{align}
\big\langle \Phi^{\text{free}}_{p,m} (z,\bar{z})\big\rangle^{\!A}_{\!R,\psi} &=
\sqrt{\frac{p}{2\pi}} \int d\varphi\ e^{im\varphi} \langle
\Phi^{\text{free}}_{pe^{i\varphi}} (z,\bar{z})\rangle^{A}_{R,\psi} \\
&= \frac{1}{\sqrt{2\pi p}} e^{im\psi} \cdot \left\{\begin{array}{ll}
\cos Rp & \text{for $m$ even}\\[4pt]
i \sin Rp & \text{for $m$ odd.}
\end{array} \right.
\label{ffonepointA}
\end{align}

In the minimal models, A-type boundary conditions are obtained using
the standard Cardy construction~\cite{Cardy:1989ir}. They are labelled by
integers $(L,M,S)$, where $0\leq L \leq k$, $M$ is $2k+4$-periodic,
$S\in \{-1,0,1,2\}$, and $L+M+S$ is even. In the geometric description~\eqref{mmgeometry}
of~\cite{Maldacena:2001ky}, these boundary conditions correspond to
branes that are straight lines
characterised by the equation
\begin{equation}
\rho \cos (\varphi -\varphi_{0}) = \rho_{0} \ ,
\end{equation}
where
\begin{align}
\rho_{0} &= \cos \frac{\pi (L+1)}{k+2}  & 
\varphi_{0} & = \frac{\pi M}{k+2} \ . 
\end{align}
Note that boundary labels $(L,M,S)$ and $(k-L,M+k+2,S+2)$ describe the
same boundary conditions, and we can always choose $L\leq k/2$ such
that the above defined~$\rho_{0}$ is positive. In the geometric
picture $(\rho_{0},\varphi_{0})$ are the coordinates of the point on
the brane that is closest to the origin. For large $k$ the distance to
the origin is given by $\rho_{0}'=\sqrt{k+2}\rho_{0}$. To make contact
with the free field theory description we want this distance
to approach the constant $R$,
\begin{equation}
\sqrt{k+2}\cos \frac{\pi (L+1)}{k+2} \to R \ .
\end{equation}
We can achieve this by scaling the boundary label as 
\begin{equation}\label{choiceforL}
L = \frac{1}{2} (k+2) - \frac{R}{\pi}\sqrt{k+2} + \mathcal{O} (1)  \ .
\end{equation}
Similarly we scale the boundary label $M$ such that the corresponding
angle $\varphi_{0}$ is constant in the limit,
\begin{equation}\label{choiceforM}
M = \frac{k+2}{\pi} \varphi_{0} + \mathcal{O} (1) \ .
\end{equation}
We expect $\varphi_{0}$ to coincide with the angle $\psi$ up to a
possible additive shift. 

The one-point function of a Neveu-Schwarz primary field $\phi_{l,m}$
for a boundary condition $(L,M,S)$ is given by (see\footnote{The sign
$(-1)^{\frac{l-m}{2}}$ comes from our field redefinition
in~\eqref{conventionchange}.} e.g.\ \cite{Maldacena:2001ky})
\begin{equation}
\langle \phi_{l,m} (z,\bar{z})\rangle^{A}_{(L,M,S)} 
= \frac{(-1)^{\frac{l-m}{2}}}{\sqrt{k+2}} \frac{\sin \frac{\pi (l+1)
(L+1)}{k+2}}{\sqrt{\sin \frac{\pi (l+1)}{k+2}}} e^{\pi i
\frac{Mm}{k+2}} \frac{1}{|z-\bar{z}|^{2h_{l,m}}} \ .
\end{equation}
For $L$ and $M$ as in~\eqref{choiceforL} and~\eqref{choiceforM}, this
behaves as
\begin{equation}\label{Atypeonepoint}
\langle \phi_{l,m} (z,\bar{z})\rangle^{A}_{(L,M,S)} 
= \frac{(k+2)^{-1/4}}{2\sqrt{\pi p}}
\left(e^{iR\frac{l+1}{\sqrt{k+2}}} - e^{i\pi (l+1) -iR\frac{l+1}{\sqrt{k+2}}}
  \right)
e^{i (\varphi_{0}-\frac{\pi}{2})m}
 \frac{1}{|z-\bar{z}|^{2h_{l,m}}} \ .
\end{equation}
To obtain the one-point function for the limit field $\Phi_{p,m}$ we
take expression~\eqref{Atypeonepoint}, multiply it by
$\alpha(k)\beta(k)$ given in~\eqref{alphabeta} and take the limit $k\to \infty$ 
while we keep $m$ constant and scale $l\approx p\sqrt{k+2}$. We arrive
at the result
\begin{equation}
\big\langle \Phi_{p,m}\big\rangle^{\!A}_{\!R,\varphi_{0}} = \frac{1}{\sqrt{2\pi p}}
e^{i (\varphi_{0}-\frac{\pi}{2})m} \cdot \left\{\begin{array}{ll}
\cos Rp & \text{for $m$ even}\\
i\sin Rp & \text{for $m$ odd}\ ,
\end{array} \right.
\end{equation}
which precisely matches the free field theory
result~\eqref{ffonepointA} upon identifying $\psi =
\varphi_{0}-\frac{\pi}{2}$.

\subsection{B-type boundary conditions}
\label{sec:freeB-type}

B-type boundary conditions in minimal models are labelled by two
integers $(L,S)$ where $0\leq L \leq k$ and $S=0,1$. The one-point
functions of Neveu-Schwarz primaries are given by (see\footnote{Note
that the sign $(-1)^{\frac{l-m}{2}}$ that one expects from the field
redefinition~\eqref{conventionchange} is absorbed by a sign hidden
inside the definition of the B-type Ishibashi states
in~\cite{Maldacena:2001ky}.} e.g.\ \cite{Maldacena:2001ky})
\begin{equation}
\langle \phi_{l,m} (z,\bar{z})\rangle^{B}_{(L,S)} =
\sqrt{2}\,\frac{\sin \frac{\pi (l+1) (L+1)}{k+2}}{\sqrt{\sin \frac{\pi
(l+1)}{k+2}}} \delta_{m,0} |z-\bar{z}|^{-2h_{l,m}} \ .
\end{equation}
Geometrically these correspond to two-dimensional
discs~\cite{Maldacena:2001ky} where the coordinate of the boundary is
given by $\rho_{1}=\sin \frac{\pi (L+1)}{k+2}$. We expect that we can
define two limits: one for which the disc shrinks to a point to
describe a zero-dimensional brane in the free theory, and one for which
the disc covers the whole plane corresponding to a two-dimensional
brane in the free theory.

Let us first consider the zero-dimensional brane. We keep the label $L$
fixed, such that the radius of the disc, $\rho_{1}'=\sqrt{k+2}\sin
\frac{\pi (L+1)}{k+2}$, goes to zero. One readily obtains the
corresponding one-point function
\begin{equation}
\big\langle \Phi_{p,m} (z,\bar{z})\big\rangle^{\!B}_{\!(L,S)} = 
\sqrt{\pi p} (L+1)\delta_{m,0} |z-\bar{z}|^{-2h_{p}} \ ,
\end{equation}
which is an integer multiple of the one-point function for $L=0$, so
it describes a stack of $L+1$ elementary branes. This is related to the
fact that in minimal models the B-type boundary conditions with $L>0$ can
be obtained from a superposition of boundary conditions with $L=0$ by
a boundary renormalisation group flow that becomes short when
$k\to\infty$~\cite{Fredenhagen:2003xf}.

In the free theory, for a zero-dimensional brane at the
origin, the one-point function of Neveu-Schwarz primary fields
$\Phi_{p}^{\text{free}}$ is simply\footnote{Note that the
zero-dimensional brane cannot couple to the R-R sector, because we
chose the projection such that the one-dimensional brane
couples to it. Therefore the prefactor is simply $2^{-D/4}=2^{-1/2}$
(compare the discussion in footnote~\ref{fn:branes} on
page~\pageref{fn:branes}).}
\begin{equation}
\big\langle \Phi_{\mathbf{p}}^{\text{free}}
(z,\bar{z})\big\rangle^{\!B}_{\!(0)} =
\frac{1}{\sqrt{2}} |z-\bar{z}|^{-2h_{p}} \ ,
\end{equation}
which in the radial basis reads 
\begin{equation}
\big\langle \Phi_{p,m}^{\text{free}} (z,\bar{z})\big\rangle^{\!B}_{\!(0)} = 
\sqrt{\pi p} \,\delta_{m,0} \frac{1}{|z-\bar{z}|^{2h_{p}}} \ ,  
\end{equation}
in precise agreement with the minimal model computation.
\smallskip

\label{pg:electric}
On the other hand, we can look at two-dimensional branes. There is a
one-parameter family of those that differ in the strength of a
constant electric background field. The electric field can be labelled
by an angle\footnote{where $\sin \phi =\frac{2f}{1+f^{2}}$ and $\cos
\phi =\frac{1-f^{2}}{1+f^{2}}$ for an electric field strength
$F_{\mu\nu}=\begin{pmatrix}0 & f\\ -f & 0 \end{pmatrix}$}
\mbox{$-\pi<\phi<\pi$} (see e.g.\
\cite{Abouelsaood:1986gd,DiVecchia:1999fx} and the discussion
in~\cite{Gaberdiel:2004nv}). The boundary conditions are characterised
by the one-point functions
\begin{equation}
\big\langle \Phi_{\mathbf{p}}^{\text{free}}
(z,\bar{z})\big\rangle^{\!B}_{\!\phi} = 
\frac{1}{\sqrt{2}\cos \frac{\phi}{2}}\delta^{(2)} (\mathbf{p}) \ .
\end{equation}
Instead of working with the delta distribution directly, it is more
convenient to apply it on a test function $\zeta (\mathbf{p})$, i.e.\
we look at a smeared one-point function
\begin{equation}
\Big\langle \int d^{2}p\ \zeta (\mathbf{p})\Phi_{\mathbf{p}}^{\text{free}}
(z,\bar{z}) \Big\rangle^{\!\!B}_{\!\!\phi} = \frac{1}{\sqrt{2}\cos \frac{\phi}{2}}\zeta (0) \ .
\end{equation}
For a comparison to the minimal model limit, we express it in terms
of the radial basis,
\begin{align}
\Big\langle \int dp\ \sum_{m} \zeta_{p,-m} \Phi_{p,m}^{\text{free}}
(z,\bar{z})\Big\rangle^{\!\!B}_{\!\!\phi}
=\Big\langle \int d^{2}p\ \zeta (\mathbf{p})\Phi_{\mathbf{p}}^{\text{free}}
(z,\bar{z}) \Big\rangle^{\!\!B}_{\!\!\phi} &= \frac{1}{\sqrt{2}\cos \frac{\phi}{2}}\zeta (0) \\
&=
\frac{1}{\sqrt{2}\cos \frac{\phi}{2}}\frac{\zeta_{p,0}}{\sqrt{2\pi p}}\bigg|_{p=0}\ ,
\end{align}
where 
\begin{equation}
\zeta_{p,m} = \sqrt{\frac{p}{2\pi}} \int d\varphi\ e^{im\varphi} \, \zeta
(pe^{i\varphi}) \ .
\end{equation}
We can reformulate this as 
\begin{subequations}
\begin{align}
\big\langle \Phi_{p,m}^{\text{free}} (z,\bar{z})\big\rangle^{\!B}_{\!\phi} &= 0 \qquad
\text{for}\ m\not= 0\\
\Big\langle \sqrt{2\pi}\int_{0}^{\infty}dp\ \sqrt{p}\,\chi
(p)\Phi_{p,0}^{\text{free}} (z,\bar{z})\Big\rangle^{\!\!B}_{\!\!\phi} &= \frac{1}{\sqrt{2}\cos \frac{\phi}{2}}\chi (0) \ ,
\label{D2free}
\end{align}
\end{subequations}
for suitable test functions $\chi$ on the positive real line.

We expect to get these boundary conditions from the minimal models by
considering B-type boundary conditions that correspond to a disc
covering the whole two-dimensional space in the minimal model
geometry. These are labelled by $(L,S)$ where $L$ is scaled linearly
with $k$, $L=\lfloor \Lambda (k+2)\rfloor$. The minimal model one-point functions behave as
\begin{equation}
\langle \phi_{l,m} (z,\bar{z})\rangle^{B}_{(\lfloor \Lambda
(k+2)\rfloor,S)} 
\approx  \sqrt{\frac{2 (k+2)}{\pi (l+1)}}\, \sin \left(\pi \Lambda (l+1) \right) \delta_{m,0} |z-\bar{z}|^{-2h_{l,m}} \ .
\end{equation}
The sine function in the numerator oscillates
rapidly as a function of $l$. Therefore the one-point function of
$\Phi_{p,0}$ is suppressed for non-zero $p$ as expected. To evaluate
the contribution at $p=0$, we consider the one-point function for
fields smeared by a test function $\chi$,
\begin{align}
&\Big\langle \sqrt{2\pi}\int_{0}^{\infty} dp\ \sqrt{p}\,\chi (p)\,
\Phi_{p,0} (z,\bar{z})\Big\rangle^{\!\!B}_{\!\!(\lfloor \Lambda
(k+2)\rfloor,S)}\nonumber\\
&\qquad = \lim_{k\to\infty} \sqrt{2\pi}\,\sqrt{2}\,(k+2)^{-1/4} \sum_{l\
\text{even}} \left(\frac{l+1}{\sqrt{k+2}}\right)^{\frac{1}{2}}\, \chi
\left(\frac{l+1}{\sqrt{k+2}} \right)
\langle \phi_{l,0} (z,\bar{z})\rangle^{B}_{(\lfloor \Lambda
(k+2)\rfloor,S)} \nonumber\\
&\qquad = \lim_{k\to\infty}  2 \sqrt{2}\sum_{l\ \text{even}} \sin
\left(\pi\Lambda (l+1) \right) \,\chi
\left(\frac{l+1}{\sqrt{k+2}} \right) |z-\bar{z}|^{-2h_{l,0}} \nonumber\\
&\qquad = \frac{\sqrt{2}}{\sin \pi \Lambda}\chi (0) \ ,
\end{align}
which equals twice the result in eq.\ \eqref{D2free} if we set 
\begin{equation}\label{identificationofphi}
\phi=\pm 2\pi\big(\Lambda-\tfrac{1}{2}\big)\ .
\end{equation}
Therefore the limiting boundary condition is not elementary, but a
superposition of two two-dimensional branes in the free theory. A closer analysis (e.g.\
by looking at the relative spectrum to the zero-dimensional brane)
reveals that in fact it is a superposition of two branes with opposite
electric field (corresponding to the two possible signs of
$\phi$ in~\eqref{identificationofphi}). This is in accordance with the identification of B-type
boundary states in minimal models under $L\leftrightarrow k-L$, which amounts to
the identification $\Lambda\leftrightarrow 1-\Lambda$ corresponding to
a switch of the sign in~\eqref{identificationofphi}.
 
This concludes our discussion of the free field limit, and we turn now
to the continuous orbifold limit.

\section{Continuous orbifold limit}
\label{sec:contorbi}

In~\cite{Fredenhagen:2012rb} we constructed a limit of minimal models
where both field labels $l$ and $m$ are sent to infinity such that both
the conformal weight and the $U(1)$ charge are kept fixed. The
resulting theory contains a spectrum of primary fields that is
continuous in the $U(1)$ charge. In this section we want to interpret
this limit as a continuous orbifold of a free theory, where the $U(1)$
charge serves as a twist parameter.
  
The possibility to construct continuous orbifolds by gauging a
continuous global symmetry group was recently explored
in~\cite{Gaberdiel:2011aa} where the non-Abelian orbifold 
$SU(2)_{1}/SO(3)$  was analysed. The theory we want to consider is the $N= (2,2)$
supersymmetric theory of two uncompactified bosons and two fermions
orbifolded by the rotation group $SO (2)\simeq U (1)$. 

\subsection{The orbifold}

Notations and conventions follow closely the ones
in~\cite{Gaberdiel:2004nv}.  We start by defining the real bosonic
coordinates $X^1(z,\bar z),X^2(z,\bar z)$ and their fermionic
counterparts $\psi^1(z,\bar z),\psi^2(z,\bar z)$. We rearrange the
fields to work on the complex plane with one free complex fermion,
namely defining
\begin{subequations}
\begin{align}
\phi &=\tfrac{1}{\sqrt 2}(X^1+iX^2) & \phi^* &=\tfrac{1}{\sqrt 2}(X^1-iX^2)\\[4pt]
\psi &=\tfrac{1}{\sqrt 2}(\psi^1+i\psi^2) & \psi^* &=\tfrac{1}{\sqrt 2}(\psi^1-i\psi^2)
\ ,
\end{align}
\end{subequations}
such that the mode expansion of the (holomorphic) fields reads
\begin{subequations}
\begin{align}
\de\phi &=-i\sum_{\substack{m\in\Z}}\alpha_m z^{-m-1} & \de\phi^* &=-i\sum_{\substack{m\in\Z}}\alpha^*_m z^{-m-1}\\
\psi &=\sum_{\substack{r\in\Z+\eta}}\psi_r z^{-r-\frac{1}{2}} & \psi^*
&=\sum_{\substack{r\in\Z+\eta}}\psi^*_r z^{-r-\frac{1}{2}}\ ,
\end{align}
\end{subequations}
where $\eta=0,\frac12$ in the Ramond and Neveu-Schwarz sector
respectively. The antiholomorphic case is analogous. For simplicity we
will restrict the following discussion to the Neveu-Schwarz
sector. The modes respect the algebra of one free complex boson and
one free Neveu-Schwarz complex fermion:
\begin{subequations}
\begin{align}  
  [\alpha_m, \alpha^*_n] &= m\,\delta_{m,-n}  &  \{\psi_r,\psi^*_s\} &= \delta_{r,-s}\\[4pt] 
  [\alpha_m,\alpha_n] &= [\alpha^*_m, \alpha^*_n] = 0 & \{\psi_r,\psi_s\}&=\{\psi^*_r,\psi^*_s\}=0 \ .
\end{align}
\end{subequations}
We can explicitly realise the $N =2$ superconformal algebra by
defining the generators through our holomorphic fields as
\begin{subequations}\label{SUSY-generators}
\begin{align}
T&=-\de\phi\de\phi^*-\frac12(\psi^*\de\psi+\psi\de\psi^*) & J&=-\psi^*\psi\\
G^+&=i\sqrt 2\,\psi\de\phi^* & G^-&=i\sqrt 2\,\psi^*\de\phi \ ,
\end{align}
\end{subequations}
and similarly for their antiholomorphic counterparts.

We want to end up with an $N=(2,2)$ theory; we therefore choose the
action of the orbifold group in such a way that the currents
in~\eqref{SUSY-generators} are invariant under the transformation and
supersymmetry is not broken. In particular we choose the $U(1)$ action
on the fields as follows
\begin{subequations}
\begin{align}
U(\theta)\cdot\phi&=e^{i\theta}\phi & U(\theta)\cdot\phi^*&=e^{-i\theta}\phi^* \\[4pt]
U(\theta)\cdot\psi&=e^{i\theta}\psi & U(\theta)\cdot\psi^*&=e^{-i\theta}\psi^*\ ,
\end{align}
\end{subequations}
so that in terms of the coordinates $X^1,X^2$ on the plane it is realised by the rotation matrix
\begin{align}
U(\theta)\cdot\vec X\equiv\mathcal{R}_{\theta}\cdot \vec X=\left(
\begin{array}{cc}
\cos{\theta}&-\sin{\theta}\\
\sin{\theta}& \cos{\theta}
\end{array}
\right)\cdot\left(
\begin{array}{c}
X^1\\
X^2
\end{array}
\right)\ .
\end{align}
The action of the group on the field modes is thus
\begin{subequations}
\begin{align}
\alpha_n&\mapsto e^{i\theta}\alpha_n & \alpha^*_n&\mapsto e^{-i\theta}\alpha^*_n\\[4pt]
\psi_r&\mapsto e^{i\theta}\psi_r & \psi^*_r&\mapsto e^{-i\theta}\psi^*_r\ .
\end{align}
\end{subequations}

\subsection{Partition function}

We now want to determine the partition function of the orbifold. We
first look at the Neveu-Schwarz part, and work with the full
supersymmetric Hilbert space. To compare with the minimal models
we will later perform a (GSO-like) projection by $\frac{1}{2} (1+
(-1)^{F+\bar{F}})$ onto states of even fermion number.

By inserting a twist operator we obtain the $\theta$-twined characters
\begin{equation}\label{th0-comp}
\torus{\theta}{0}\,=
\Tr_{\cH_{\text{free}}^{\text{NS}}}\left(
U(\theta)q^{L_0-\frac18}\bar{q}^{\bar{L}_0-\frac18}\right) \ ,
\end{equation}
where we denoted by $\mathcal{H}_{\text{free}}^{\text{NS}}$ the (unprojected)
Neveu-Schwarz part of the Hilbert space of the free
theory.

The orbifold group acts non-trivially on the vacua labelled
by the momentum on the plane,
\begin{equation}
|\vec{p}\ \rangle\ \longmapsto\ |\mathcal{R}_{\theta}\cdot\vec{p}\ \rangle\ ,
\end{equation}
so that the momentum dependent part of equation~\eqref{th0-comp} becomes
\begin{equation}\label{int-zero-modes}
\int d^2p\  \delta^2(\mathcal{R}_{\theta}\cdot \vec p-\vec p)\ (q\bar{q})^{\frac{|\vec{p}|^2}{4}}=\int d^2p\
\frac{1}{\det({\mathcal{R}_{\theta}-1)}} \delta^2(\vec p)\ (q\bar{q})^{\frac{|\vec{p}|^2}{4}} \ .
\end{equation}
The $\theta$-twined character is then
\begin{align}\label{th0-c}
\torus{\theta}{0}\,&=
\Tr_{\mathcal{H}_{\text{free}}^{\text{NS}}}\left( U(\theta)q^{L_0-\frac18}\bar{q}^{\bar{L}_0-\frac18}\right)\nonumber\\
&=\int d^2p\  \frac{\delta^2(\vec p)}{\det({\mathcal{R}_{\theta}-1)}}\ (q\bar q)^{\frac{|\vec{p}|^2}{4}}
\left|q^{-\frac18}\prod_{n=0}^{\infty}\frac{(1+q^{n+\frac12}e^{i\theta})(1+q^{n+\frac12}e^{-i\theta})}{(1-q^{n+1}e^{i\theta})(1-q^{n+1}e^{-i\theta})}\right|^2
\nonumber\\
&=\left|\frac{\vartheta_3(\tau,\frac{\theta}{2\pi})}{\vartheta_1(\tau,\frac{\theta}{2\pi})}\right|^2
\ .
\end{align}
We then act with a modular S-transformation on the complex modulus of
the torus ($\tau\mapsto -\frac{1}{\tau}$) to get from the $\theta$-twined free
character to the character of the $\theta$-twisted sector,
\begin{align}
\torus {\theta}{0}\quad\overset{S}{\longmapsto}\quad\torus{0}{\theta}\ .
\end{align}
We can benefit from known transformation properties of the $\vartheta$-functions, in particular
\begin{equation}\label{modular-thetas}
\frac{\vartheta_3(-\frac{1}{\tau},\nu)}{\vartheta_1(-\frac{1}{\tau},\nu)}=i\frac{\vartheta_3(\tau,\nu\tau)}{\vartheta_1(\tau,\nu\tau)}\ ,
\end{equation}
so that the $\theta$-twisted sector reads
\begin{align}
\torus{0}{\theta}\,=&\,
\Tr_{\mathcal{H}_{\theta}^{\text{NS}}}\left( q^{L_0-\frac18}\bar{q}^{\bar{L}_0-\frac18}\right)
=\left|\frac{\vartheta_3(\tau,\frac{\tau\theta}{2\pi})}{\vartheta_1(\tau,\frac{\tau\theta}{2\pi})}\right|^2\\
=&\left|q^{-\frac18+\frac{\theta}{4\pi}}
\prod_{n=0}^{\infty}\frac{(1+q^{n+\frac12+\frac{\theta}{2\pi}})(1+q^{n+\frac12-\frac{\theta}{2\pi}})}{(1-q^{n+\frac{\theta}{2\pi}})(1-q^{n+1-\frac{\theta}{2\pi}})}\right|^2
\ .
\end{align}
We can now get the $\theta'$-twined character over the
$\theta$-twisted sector by acting once more with the orbifold group
on the modes. We get the following:
\begin{align}\label{thpth}
\torus{\theta'}{\theta}\,
&=\Tr_{\mathcal{H}_{\theta}^{\text{NS}}}\left( U(\theta')q^{L_0-\frac18}\bar{q}^{\bar{L}_0-\frac18}\right)\nonumber\\
&=\left|q^{-\frac18+\frac{\theta}{4\pi}}
\prod_{n=0}^{\infty}\frac{(1+q^{n+\frac12+\frac{\theta}{2\pi}}e^{i\theta'})(1+q^{n+\frac12-\frac{\theta}{2\pi}}e^{-i\theta'})}{(1-q^{n+\frac{\theta}{2\pi}}e^{i\theta'})(1-q^{n+1-\frac{\theta}{2\pi}}e^{-i\theta'})}\right|^2\nonumber\\
&=\left|\frac{\vartheta_3(\tau,\frac{\tau\theta+\theta'}{2\pi})}{\vartheta_1(\tau,\frac{\tau\theta+\theta'}{2\pi})}\right|^2\ ,
\end{align}
which is the expression we are interested in.

The contribution of a $\theta$-twisted sector to the unprojected
partition function is
therefore obtained by integrating equation~\eqref{thpth} over the twisting parameter~$\theta'$,
\begin{align}
\mathcal{P}_{\theta-\text{twisted}}^{\text{NS}}&=\frac{1}{2\pi}\int_{0}^{2\pi}d\theta'\,\torus{\theta'}{\theta}= \int_{0}^{2\pi}\frac{d\theta'}{2\pi}\ \Tr_{\mathcal{H}_{\theta}^{\text{NS}}}\left(U(\theta')q^{L_0-\frac18}\bar{q}^{\bar{L}_0-\frac18} \right)\\
&=\int_{0}^{2\pi}\frac{d\theta'}{2\pi}\left|\frac{\vartheta_3(\tau,\frac{\tau\theta+\theta'}{2\pi})}{\vartheta_1(\tau,\frac{\tau\theta+\theta'}{2\pi})}\right|^2\ .
\label{Z-twisted-implicit}
\end{align}
Using some identities of appendix~C in~\cite{Gaberdiel:2004nv} the modular functions can be recast in the form
\begin{equation}\label{th-identity}
\frac{\vartheta_3(\tau,\nu)}{\vartheta_1(\tau,\nu)}=-2i\frac{\vartheta_3(\tau,0)}{\eta^3(\tau)}\sum_{n=0}^{\infty}\cos{\left[2\pi(n+1/2)(\nu-\tau/2)\right]}\ \frac{q^{\frac n2+\frac14}}{1+q^{n+\frac12}}\ ,
\end{equation}
so that the integral~\eqref{Z-twisted-implicit} becomes
\begin{equation}\label{Z-twisted-explicit}
\mathcal{P}_{\theta-\text{twisted}}^{\text{NS}}=4\left|\frac{\vartheta_3(\tau,0)}{\eta^3(\tau)}\right|^2\sum_{n,\bar{n}=0}^{\infty}
\frac{q^{\frac n2+\frac14}\bar{q}^{\frac {\bar n}{2}+\frac14}}{(1+q^{n+\frac12})(1+\bar{q}^{\bar n+\frac12})}I^{\theta}_{n,\bar{n}}
\end{equation}
with
\begin{align}
I^{\theta}_{n,\bar{n}}&=\int_0^{2\pi} \frac{d\theta'}{2\pi}\cos\left[(n+\tfrac{1}{2}) (\tau(\theta-\pi)+\theta') \right]\, \cos\left[ (\bar n+\tfrac{1}{2}) (\bar{\tau}(\theta-\pi)+\theta')\right]\\
&=\frac{\delta_{n,\bar n}}{2}\
\cos\left[ (n+\tfrac{1}{2})(\pi-\theta)(\tau-\bar{\tau})\right]\ .
\label{thp-integral}
\end{align}
Inserting~\eqref{thp-integral} into~\eqref{Z-twisted-explicit},
evaluating the sum over $\bar{n}$, and combining the cosine with
the~$q,\bar q$~dependent part of the numerator, we arrive at
\begin{equation}
\mathcal{P}_{\theta-\text{twisted}}^{\text{NS}}=\left|\frac{\vartheta_3(\tau,0)}{\eta^3(\tau)}\right|^2
\sum_{n=0}^{\infty}\frac{q^{\frac{\theta}{2\pi}(n+\frac12)}\bar{q}^{\frac{\theta}{2\pi}(n+\frac12)}+q^{(1-\frac{\theta}{2\pi})(n+\frac12)}\bar{q}^{(1-\frac{\theta}{2\pi})(n+\frac12)}}
{(1+q^{n+\frac12})(1+\bar{q}^{n+\frac12})}\ .
\end{equation}
The unprojected supersymmetric partition function is then obtained by integrating over all twisted sectors 
\begin{align}\label{cont-orb-Z-infty}
\mathcal{P}^{\text{NS}}_{\mathbb{C}/U(1)}&=\int_0^{2\pi}\frac{d\theta}{2\pi}
\mathcal{P}_{\theta-\text{twisted}}^{\text{NS}}=
\sum_{n=0}^{\infty}\int_{-2\pi}^{2\pi}\frac{d\theta}{2\pi}\frac{q^{\frac{|\theta|}{2\pi}(n+\frac12)}\bar{q}^{\frac{|\theta|}{2\pi}(n+\frac12)}}{(1+q^{n+\frac12})(1+\bar{q}^{n+\frac12})}\nonumber\\
&=\left|\frac{\vartheta_3(\tau,0)}{\eta^3(\tau)}\right|^2\sum_{n=0}^{\infty}\left|\frac{1}{1+q^{n+\frac12}}\right|^2\int ^1_{-1}dQ \ (q\bar q)^{|Q|(n+\frac12)}\nonumber\\
&=\sum_{n=0}^{\infty}\int_{-1}^{1}dQ\left|\chi^{I}_{|Q|(n+\frac12),Q}\right|^2\ ,
\end{align} 
where we used the definitions of appendix~\ref{app:characters} for the $c=3$ character $\chi^I$.

The last expression is ill-defined: the integration over $Q$ gives
\begin{equation}
 \mathcal{P}^{\text{NS}}_{\mathbb{C}/U(1)}=\frac{1}{2\pi\tau_2}\left|\frac{\vartheta_3(\tau,0)}{\eta^3(\tau)}\right|^2\sum_{n=0}^{\infty}\frac{1-(q\bar q)^{n+\frac12}}{\big|1+q^{n+\frac12}\big|^2}\ \frac{1}{n+\frac12}\ ,
\end{equation} 
which exhibits a logarithmic divergence when we sum over $n$. The
fields that contribute to this divergence are the chargeless ones, as
one can see by looking at the large $n$ asymptotic behaviour of the
function~\eqref{cont-orb-Z-infty}: the fraction in front of the
integral tends to one and the integrand localises
around~$Q\sim0$. Therefore a sensible regulator would screen away the
untwisted fields. We define
\begin{equation}\label{cont-orb-Z-reg}
\mathcal{P}^{\text{NS},(r)}_{\mathbb{C}/U(1)}:=\left|\frac{\vartheta_3(\tau,0)}{\eta^3(\tau)}\right|^2\sum_{n=0}^{\infty}\left|\frac{1}{1+q^{n+\frac12}}\right|^2\int ^1_{-1}dQ \ (q\bar q)^{|Q|(n+\frac12)}\left(1-e^{2\pi ir Q}\right)\ ,
\end{equation}
which corresponds to inserting $1-e^{2\pi irJ_{0}}$ in the trace, where
$J_{0}$ is the zero mode of the $U(1)$ current $J(z)$. 
We see explicitly that this cures the logarithmic divergences of the
sum in equation~\eqref{cont-orb-Z-infty} by performing the integral over the twist $Q$,
\begin{multline}
 \mathcal{P}^{\text{NS},(r)}_{\mathbb{C}/U(1)}=\left|\frac{\vartheta_3(\tau,0)}{\eta^3(\tau)}\right|^2\\
\times \sum_{n=0}^{\infty}\left|\frac{1}{1+q^{n+\frac12}}\right|^2\left[\frac{1-(q\bar
 q)^{n+\frac12}}{2\pi\tau_2(n+\frac12)}-\frac{1-e^{2\pi ir}(q\bar
 q)^{n+\frac12}}{4\pi\tau_2(n+\frac12)-2\pi ir} -\frac{1-e^{-2\pi ir}(q\bar
 q)^{n+\frac12}}{4\pi\tau_2(n+\frac12)+2\pi ir} \right]\ .
\end{multline}
The summand is suppressed by $n^{-2}$ for large $n$, and the series converges.

From equation~\eqref{cont-orb-Z-reg} it is easy to write down the
(GSO-like) projected version of the regularised partition function,
which reads in the Neveu-Schwarz sector
\begin{equation}\label{cont-orb-Z-reg-proj}
Z _{\mathbb{C}/U(1)}^{\text{NS},(r)}=\frac12\left(
\left|\frac{\vartheta_3(\tau,0)}{\eta^3(\tau)}\right|^2+\left|\frac{\vartheta_4(\tau,0)}{\eta^3(\tau)}\right|^2\right)
\sum_{n=0}^{\infty}\left|\frac{1}{1+q^{n+\frac12}}\right|^2\int ^1_{-1}dQ \ (q\bar q)^{|Q|(n+\frac12)}\left(1-e^{2\pi ir Q}\right)\ .
\end{equation}

\subsubsection*{Comparison with the limit of minimal models} 
 
We now want to show that the partition function of minimal models
reproduces the result of equation~\eqref{cont-orb-Z-reg-proj} in the
limit we analysed in reference~\cite{Fredenhagen:2012rb}.  We are
thus interested in the behaviour of minimal models in the regime in
which $n=\frac{l-|m|}{2}$ is a fixed non-negative integer, and $|m|$ scales
with $k$.  The Neveu-Schwarz contribution to the partition function for the
$A_{k+2}$ minimal model reads (see appendix~\ref{app:characters} for
notations and details)
\begin{equation}\label{type-0-MM-Z}
Z^{\text{NS}}_{k} (\tau ,\nu) =
\frac12\sum_{l=0}^k\sum_{\substack{m=-l\\ m+l\ \text{even}}}^l  \left[
\chi^{\text{NS}}_{l,m}\bar
\chi^{\text{NS}}_{l,m}(q,z)+\chi^{\text{NS}}_{l,m}\bar \chi^{\text{NS}}_{l,m}(q,-z)\right]\ ,
\end{equation}
where $z=e^{2\pi i\nu}$.
As before we first analyse the partition function before taking the
(GSO-like) projection, i.e.\ the corresponding trace is taken over the
full supersymmetric Hilbert space, 
\begin{equation}\label{redundant-Z-MM}
\begin{split} 
\mathcal{P}^{\text{NS}}_{k} (\tau ,\nu) &= \sum_{l=0}^{k}\sum_{\substack{m=-l\\ m+l\ \text{even}}}^l \chi^{\text{NS}}_{l,m}\bar \chi^{\text{NS}}_{l,m}(q,z)\\
&=\left|\frac{\vartheta_3(\tau,\nu)}{\eta^3(\tau)}\right|^2\sum_{l=0}^{k}\sum_{m=-l}^l\left|q^{\frac{(l+1)^{2}-m^2}{4(k+2)}}
\Gamma^{(k)}_{lm}(\tau,\nu)\right|^2\ .
\end{split}
\end{equation}
For large level $k$ we expect the same kind of divergence as for
the partition function of the continuous orbifold due to the almost
chargeless field. Similarly to our strategy there we insert the factor
$(1-e^{2\pi irJ_{0}})$ in the trace, and arrive at (we set $\nu =0$ in
the following)
\begin{equation}\label{PintermsofI}
\mathcal{P}^{\text{NS},(r)}_{k} (\tau) =\left|\frac{\vartheta_3(\tau,0)}{\eta^3(\tau)}\right|^2
\left[
\sum_{n=0}^{\lfloor\frac k2\rfloor}\mathcal I^{(r)}_{k,n}
\right]
\end{equation}
with
\begin{equation}\label{reg-MM}
\mathcal{I}^{(r)}_{k,n}:=2\sum_{m=1}^{k-2n}(q\bar
q)^{\frac{1}{k+2}\left(n+\frac{1}{2}\right)^{2}+\frac{m}{k+2}\left(n+\frac12\right)}
\left| \Gamma^{(k)}_{m+2n,m} (\tau ,0)\right|^2
\left(1-\cos \left( 2\pi r \tfrac{m}{k+2}\right)\right)\ .
\end{equation}
For large level $k$, the main contribution comes from
small $n$ and large $m$: the regularisation factor $(1-\cos (\cdot))$
is small unless $m$ is of order $k$, while the exponent containing the
conformal weight tells us that for large $m$ only small values of $n$
contribute significantly. 
In this limit, only one singular vector survives in $\Gamma^{(k)}_{lm}$ (the one
present in the $c=3$ representations of type $I$ in
appendix~\ref{app:characters}).  
Using the Euler-MacLaurin formula to convert the sum over $m$ into an
integral, we obtain
\begin{align}
\mathcal{I}^{(r)}_{k,n} &\approx 2\sum_{m=1}^{k-2n}(q\bar
q)^{\frac{m}{k+2}\left(n+\frac12\right)}
\left| \frac{1-q^{m+2n+1}}{(1+q^{n+\frac12})(1+q^{m+n+\frac12})}\right|^2
\left(1-\cos \left( 2\pi r \frac{m}{k+2}\right)\right) \\
&\approx 2 (k+2) \int_{0}^{1} dQ\, (q\bar q)^{-Q\left(n+\frac12\right)} 
\left| \frac{1}{(1+q^{n+\frac12})}\right|^2 
\left(1-\cos \left( 2\pi r Q \right)\right)\ .
\end{align}
Inserting this into~\eqref{PintermsofI} and comparing to~\eqref{cont-orb-Z-reg} we find
\begin{equation}
\lim_{k\to\infty} \frac{1}{k+2} \mathcal{P}^{\text{NS},(r)}_{k} (\tau) 
= \mathcal{P}^{\text{NS},(r)}_{\mathbb{C}/U(1)} (\tau) \ .
\end{equation}
The rescaling by a factor $1/(k+2)$ can be understood as follows: for
a fixed $n$ and a given small interval $[Q,Q+\Delta Q]$ there are
roughly $(k+2)\Delta Q$ ground states in the $k^{\text{th}}$ minimal
model that contribute with approximately the same weight
$(q\bar{q})^{|Q|(n+\frac{1}{2})}$. The rescaling thus corresponds to a 
rescaling of the density of states to $1$ per unit interval $\Delta Q$.
An analogous relation holds for the true (projected) partition
functions, so that indeed we recover the continuous orbifold partition
function in the limit.

\subsection{Boundary conditions}

The technology to study boundary conditions on discrete orbifold models is well
developed (see e.g.\ \cite{Billo:2000yb} and references
therein), and essentially they are also
applicable for the continuous orbifold we are considering (see also~\cite{Gaberdiel:2011aa}). 

For continuous orbifolds one meets the phenomenon that the untwisted
fields are in a sense outnumbered by the twisted fields -- in the
partition function~\eqref{cont-orb-Z-reg-proj} the untwisted,
chargeless fields give a contribution of measure zero. Therefore the
only interesting boundary conditions are those that couple to the
twisted sectors, i.e.\ fractional boundary states. To obtain those we
have to start from boundary conditions in the plane 
that are invariant under the action of the orbifold group. In our
case, these are the boundary conditions corresponding to a point-like
brane at the origin of the plane, and the boundary conditions
corresponding to space-filling branes.

Let us focus on the point-like brane. The fractional boundary
conditions are then labelled by representations of the orbifold group
$U(1)$, i.e.\ by an integer $m$. The relative spectrum for two such
boundary conditions labelled by $m$ and $m'$ follows from the usual
orbifold rules,
\begin{equation}\label{relopenstring}
\mathcal P_{m,m'}(\tilde q)=\int_{0}^{2\pi}\frac{d\theta}{2\pi}\ \chi_m(\theta)\chi^{*}_{m'}(\theta)\ \Tr_{\mathcal{H}^{\text{open}}}\left[U(\theta)\tilde q^{L_0-\frac18}\right]\ ,
\end{equation}  
where $\tilde q=e^{2\pi i\tilde \tau}$ and $\chi_m (\theta)=e^{im\theta}$ is a $U(1)$ group
character. $\mathcal{H}^{\text{open}}$ denotes the Hilbert space of
boundary fields for the point-like brane, which is just given by the
free Neveu-Schwarz vacuum representation. Note that depending on the
projection of the bulk spectrum, the point-like boundary condition
could couple to the Ramond-Ramond sector, in which case the boundary
spectrum would be projected by $\frac{1}{2} (1\pm (-1)^{F})$. The
unprojected spectrum will be denoted by $\mathcal{P}_{m,m'}$ as
introduced above. Evaluating~\eqref{relopenstring} we find
\begin{equation}\label{open-string-ampl}
\begin{split}
\mathcal P_{m,m'}(\tilde q)=&\int_{0}^{2\pi}\frac{d\theta}{2\pi}e^{i(m-m')\theta}\ 2\sin{\frac{\theta}{2}}\ \frac{\vartheta_3(\tilde\tau,\frac{\theta}{2\pi})}{\vartheta_1(\tilde\tau,\frac{\theta}{2\pi})}\\
=&
-4i\frac{\vartheta_3(\tilde\tau,0)}{\eta^3(\tilde\tau)}\sum_{n=0}^{\infty}\frac{q^{\frac n2+\frac14}}{1+q^{n+\frac12}}
\ \int_{0}^{2\pi}\frac{d\theta}{2\pi}\ \sin{\frac{\theta}{2}}\ e^{i(m-m') \theta}\cos{(n+\tfrac{1}{2})(\theta-\pi\tilde\tau)}\ ,
\end{split}
\end{equation}
where we have made again use of equation~\eqref{th-identity}. We can explicitly evaluate the integral,
\begin{multline}
\int_{0}^{2\pi}\frac{d\theta}{2\pi}\ \sin{\frac{\theta}{2}}\ e^{i\Delta m \theta}\cos{(n+\tfrac{1}{2})(\theta-\pi\tilde\tau)}\\
=\frac{1}{4i}\left[\tilde q^{\frac12(n+\frac12)}\left(\delta_{\Delta m,n}-\delta_{\Delta m-1,n}\right)+\tilde q^{-\frac12(n+\frac12)}\left(\delta_{-\Delta m+1,n}-\delta_{-\Delta m,n}\right)\right]\ ,
\end{multline}
where $\Delta m=m-m'$. Inserting this into~\eqref{open-string-ampl} we
find that the spectrum is given by single $N=2$ characters: in the
notations of appendix~\ref{app:characters} we obtain
\begin{subequations}
\begin{align}
\mathcal P_{m,m}(\tilde q)&=
\frac{\vartheta_3(\tilde \tau,0)}{\eta^3(\tilde \tau)}\left(\frac{1-\tilde q^{\frac12}}{1+\tilde q ^{\frac12}}\right)=\chi^{\text{vac}}_{0,0}(\tilde q)\\
\intertext{and (for $m\not=m'$)}
\mathcal P_{m,m'}(\tilde q) &=
\frac{\vartheta_3(\tilde \tau,0)}{\eta^3(\tilde \tau)}\ \tilde q^{|\Delta m|-\frac12}\left[\frac{1-\tilde q}{(1+\tilde q^{|\Delta m|-\frac12})(1+\tilde q^{|\Delta m| +\frac12})}\right]
=\chi^{III^{\pm}}_{|\Delta m|-\frac12,\,\pm 1}(\tilde q)\ ,
\end{align}
\end{subequations}
where the upper sign applies for $\Delta m>0$ and vice versa.
This result can now be compared to the limit of minimal
models. In~\cite{Fredenhagen:2012rb} two types of boundary conditions
were identified. They arise as limits of A-type boundary
conditions in minimal models, which are labelled by triples $(L,M,S)$
with the same range as labels for minimal model fields (see
appendix~\ref{app:minimalmodel} for the conventions). The first type
of boundary conditions is obtained by keeping the boundary labels
fixed while taking the limit. Only for $L=0$ one obtains
elementary boundary conditions. The label $S$ can be fixed to even
values for a fixed gluing condition for the supercurrents, and the two
remaining choices $S=0,2$ determine the overall sign of the
Ramond-Ramond couplings (thus distinguishing brane and anti-brane).
The relative spectrum for two such boundary conditions reads~\cite{Maldacena:2001ky}
\begin{equation}
Z^{(k)}_{(0,M,S),(0,M',S')}(\tilde{q})=\chi_{(0,M-M',S-S'+2)}(\tilde q)\ .
\end{equation} 
This is a projected part of the full supersymmetric character
$\chi^{\text{NS}}_{0,M-M'}$. For $M=M'$ this is the minimal model
vacuum character, which for $k\to\infty$ goes to the $c=3$ vacuum
character. For $M\not= M'$, using field identification
(see~\eqref{fieldidentification}) the labels can be brought to the
standard range, $(0,M-M')\sim (k,M-M'\mp (k+2))$, where the sign
depends on $M-M'$ being positive or negative. In the limit
$k\to\infty$ the corresponding character approaches a type $III$
character (see~\eqref{limit-III-minus} and~\eqref{limit-III-plus}),
\begin{equation}
\lim_{k\to\infty} \chi^{\text{NS}}_{0,M-M'\mp (k+2)} = 
\chi^{III^{\pm}}_{\frac{M-M'}{2}-\frac12,\pm 1}\ .
\end{equation}
The unprojected part of the boundary spectrum thus coincides with the
spectrum for the fractional boundary conditions in the continuous
orbifold upon identifying $M=2m$. On the other hand, the spectrum in
the limit of minimal models is projected. To get agreement we
therefore need that the point-like boundary conditions in the
continuous orbifold model couple to the Ramond-Ramond sector, which
specifies the necessary (GSO-like) projection in the Ramond-Ramond
sector. Note that this is precisely opposite from the projection that
we need in the free field theory limit, which is in accordance with
the T-duality that we use in the geometric interpretation of the
equivalence of a minimal model and its $\mathbb{Z}_{k+2}$ orbifold
(see the discussion at the end of section~\ref{sec:geometry}).
\smallskip

In ref.\ \cite{Fredenhagen:2012rb}, instead of the boundary spectrum, the one-point
functions have been determined. To make contact to these results, we
perform a modular transformation to get the boundary state overlap: we rewrite the boundary
partition function~\eqref{open-string-ampl} in terms of the modulus
$\tau=-\frac{1}{\tilde{\tau}}$ using the known
transformation properties~\eqref{modular-thetas},
\begin{equation}
\mathcal P_{m,m'}(\tilde{q})=
2i\int_{0}^{2\pi}\frac{d\theta}{2\pi}e^{i(m-m')\theta}\sin{\frac{\theta}{2}}\
\frac{\vartheta_3(\tau,\frac{\tau\theta}{2\pi})}{\vartheta_1(\tau,\frac{\tau\theta}{2\pi})}\
.
\end{equation}
The ratio of $\vartheta$-functions can be rewritten using \mbox{eq.\ \eqref{th-identity}},
\begin{equation}
\begin{split}
\frac{\vartheta_3(\tau,\frac{\tau\theta}{2\pi})}{\vartheta_1(\tau,\frac{\tau\theta}{2\pi})}=&-2i\frac{\vartheta_3(\tau,0)}{\eta^3(\tau)}\sum_{n=0}^{\infty}\cos{\left[2\pi(n+1/2)\left(\frac{\tau\theta}{2\pi}-\tau/2\right)\right]}\ \frac{q^{\frac n2+\frac14}}{1+q^{n+\frac12}}\\
=&-i\frac{\vartheta_3(\tau,0)}{\eta^3(\tau)}\sum_{n=0}^{\infty}
\frac{q^{(n+\frac12)\frac{\theta}{2\pi}} + q^{(n+\frac12)(1-\frac{\theta}{2\pi})}     }{1+q^{n+\frac12}}\ ,
\end{split}
\end{equation}
so that we obtain
\begin{equation}\label{overlap-contorbi}
\begin{split}
\mathcal P_{m,m'}(\tilde q)=&
\frac{\vartheta_3(\tau,0)}{\eta^3(\tau)}\int_{0}^{2\pi}\frac{d\theta}{2\pi}e^{i(m-m')\theta}\,
2\sin{\frac{\theta}{2}}\,\sum_{n=0}^{\infty}
\frac{q^{(n+\frac12)\frac{\theta}{2\pi}} + q^{(n+\frac12)(1-\frac{\theta}{2\pi})}}{1+q^{n+\frac12}}\\
=& \sum_{n=0}^{\infty}\int_{-1}^{+1}dQ\ 2\sin \left(\pi |Q| \right)
e^{2\pi i (m-m')Q}\,\chi^{I}_{|Q|(n+\frac12),Q}(q)\ .\end{split}
\end{equation}
If we do the same analysis for the projected spectrum, we find
\begin{multline}
Z_{m,m'}(\tilde{q})=\sum_{n=0}^{\infty}\int_{-1}^{+1}dQ\ \sin\left(\pi |Q|\right)
e^{2\pi
i(m-m')Q}\left(\chi^{\text{NS}}_{|Q|(n+\frac12),Q}(q)+\chi^{\text{R}}_{\frac{1}{8}+|Q|(n+1),Q}(q)\right)\\
+ \int_{-\frac{1}{2}}^{\frac{1}{2}}dQ\  \sin \left(\pi
\big|Q-\tfrac{1}{2}\big| \right) e^{2\pi i (m-m')(Q-\frac{1}{2})}\,
\chi^{\text{R}^{0}}_{\frac{1}{8},Q} (q)
\ .
\end{multline}
Comparing with the formulae presented in reference~\cite[eqs
(4.5)-(4.7)]{Fredenhagen:2012rb}, we find perfect agreement with the
one-point functions given there for the discrete A-type boundary
states of the limit theory for $L=0$ and with the identification
$M=2m$.  \medskip

Along similar lines let us briefly discuss boundary conditions that
correspond to two-dimensional branes. As we discussed at the end of
section~\ref{sec:freeB-type} on page~\pageref{pg:electric}, there is a
one-parameter family of those that differ in the strength of a
constant electric background field, which can be labelled by an angle
$\phi$. In the orbifold the boundary conditions
obtain an additional integer label $m$ that determines the
corresponding representation of $U(1)$. The unprojected part of the
annulus partition function with such a two-dimensional boundary condition
labelled by $\phi$ and $m$, and a zero-dimensional boundary condition
labelled by $m'$ is then (using again~\eqref{th-identity})
\begin{align}
\mathcal{P}_{(\phi,m),m'} (\tilde{q}) &= \int_{0}^{2\pi} \frac{d\theta}{2\pi}
e^{i (m-m'+\frac{1}{2})\theta} \, i \frac{\vartheta_{3}
(\tilde{\tau},\frac{\theta+ (\phi
+\pi)\tilde{\tau}}{2\pi})}{\vartheta_{1} (\tilde{\tau},\frac{\theta+ (\phi
+\pi)\tilde{\tau}}{2\pi})} \\
& = \frac{\vartheta_{3}(\tilde{\tau},0)}{\eta^{3}(\tilde{\tau})}
\frac{\tilde{q}^{\frac{\pi \mp \phi}{2\pi}|\Delta m
+\frac{1}{2}|}}{1+\tilde{q}^{|\Delta m+\frac{1}{2}|}} \ ,
\label{relspectrum}
\end{align}
where the upper sign corresponds to $\Delta m=m-m'\geq 0$, and the lower
one to $\Delta m<0$. These are the type $I$ characters
$\chi^{I^{\pm}}_{(n+\frac{1}{2})|Q|,Q}$ for
charge $|Q|=\frac{\pi \mp \phi}{2\pi}$ and $n=|\Delta
m+\frac{1}{2}|-\frac{1}{2}$. Note that
this is precisely the result we expect from the limit of minimal
models: in \mbox{ref.\ \cite{Fredenhagen:2012rb}} we constructed a continuous
family of A-type boundary states labelled by $Q,N$ as a limit of
minimal model boundary states with labels
\begin{equation}
(L,M,S) = (|\lfloor -Q (k+2)\rfloor| +2N, \lfloor-Q (k+2)\rfloor ,0) \ ,
\end{equation}
where $\lfloor x\rfloor$ denotes the greatest integer smaller or equal $x$.
Their relative spectrum (without projection) to a boundary
condition $(0,M',0)$ with fixed $M'$ is simply given by 
$\chi^{\text{NS}}_{L,M-M'}$, and in the limit we find (see
appendix~\ref{app:limit-char}) 
\begin{equation}
\chi^{\text{NS}}_{|\lfloor -Q (k+2)\rfloor|+2N,\lfloor -Q (k+2)\rfloor -M'}
\to \left\{ \begin{array}{ll}
\chi^{I^{+}}_{|Q|\,|N-\frac{M'}{2}+\frac{1}{2}|,Q} & Q>0,\ N\geq \frac{M'}{2} \\[6pt]
\chi^{I^{+}}_{|Q-1|\,|N-\frac{M'}{2}+\frac{1}{2}|,Q-1} & Q>0,\ N<\frac{M'}{2}\\[6pt]
\chi^{I^{-}}_{|Q|\,|N+\frac{M'}{2}+\frac{1}{2}|,Q} & Q<0,\ N\geq -\frac{M'}{2}\\[6pt]
\chi^{I^{-}}_{|Q+1|\,|N+\frac{M'}{2}+\frac{1}{2}|,Q+1} & Q<0,\ N<-\frac{M'}{2}
\end{array}\right. \ . 
\end{equation}
These are the type $I$ characters that we found above
in~\eqref{relspectrum} if we identify 
\begin{align}
\phi &=2\pi \big(-Q\pm \tfrac{1}{2}\big) & m &=-\tfrac{1}{2} \pm
\big(N+\tfrac{1}{2}\big) \ ,
\end{align}
where the upper sign applies for $Q>0$, and the lower for $Q<0$.

\section{Discussion}
\label{sec:discussion}

We have shown that one can obtain two different limits of the sequence
of $N=(2,2)$ minimal models, and we have discussed how these limits
can be understood geometrically. The first limit theory is simply a
free field theory, the second limit theory is the
non-rational theory of~\cite{Fredenhagen:2012rb}, and we have shown
that it can be described as a continuous orbifold
$\mathbb{C}/U(1)$. The latter observation is reminiscent of the
recent interpretation of the limit of Virasoro minimal models as a
continuous orbifold $SU(2)_{1}/SO(3)$ \cite{Gaberdiel:2011aa}.

It would be interesting to explore similar limits in the case of other
series of \mbox{$N=(2,2)$} superconformal models, like the Grassmannian
Kazama-Suzuki models~\cite{Kazama:1989qp} based on $SU(n+1)/U(n)$. Again one might expect
to find different possible limit theories; in fact there might be a
greater variety of limits, because in addition to the $U(1)$ charge
there are charges associated to currents of higher spin for which one
might have the freedom to scale them while taking the limit. One is
tempted to speculate that the limit theory corresponding to fixed
charges is again described by a continuous orbifold
$\mathbb{C}^{n}/U(n)$. It would be interesting to study this in
detail. This could also be of relevance in the context of the
supersymmetric generalisation of minimal model
holography~\cite{Gaberdiel:2010pz,Gaberdiel:2012uj}, where a limit of
Kazama-Suzuki models occurs in the conjectured holographic
dual of supersymmetric higher-spin theories on
three-dimensional asymptotically Anti-de Sitter
space-times~\cite{Creutzig:2011fe,Candu:2012jq,Hanaki:2012yf,Ahn:2012fz,Candu:2012tr}.

\subsection*{Acknowledgements} 
We would like to thank Matthias Gaberdiel, Ilarion Melnikov, Volker Schomerus and
Roberto Volpato for interesting and useful discussions.

\appendix
\section{Characters}\label{app:characters} 

In this appendix we collect results about characters for $N=2$
theories and their limits. A general character over a sector
$\mathcal{H}_{h,Q}$ labelled by $h,Q$ (eigenvalues of the
$L_0,J_0$~generators respectively) of the $N=2$ superconformal algebra
is defined as
\begin{equation}
\chi_{h,Q}^{\phantom {\text{NS}}}(q,z)=\Tr_{\mathcal{H}_{h,Q}}q^{L_0-\frac{c}{24}}z^{J_0}\ ,
\end{equation}
with $q=e^{2\pi i \tau}, z=e^{2\pi i \nu}$. In the main text we often
make use of the following shorthand notation for characters specialised
to $z=1$,
\begin{equation}
\chi_{h,Q}(q)\equiv\chi_{h,Q}(q,1)\ .
\end{equation}
Throughout the text we use $\vartheta$ and $\eta$ functions with the following conventions:
\begin{equation*}
\begin{split}
 \vartheta_1(\tau,\nu)&\ =-iz^{\frac12}q^{\frac18}\prod\limits_{n=0}^{\infty}(1-q^{n+1}z)(1-q^{n}z^{-1})(1-q^{n+1})\\
 \vartheta_2(\tau,\nu)&\ =z^{\frac12}q^{\frac18}\prod\limits_{n=0}^{\infty}(1+q^{n+1}z)(1+q^{n}z^{-1})(1-q^{n+1})\\
\vartheta_3(\tau,\nu)&\ =\prod\limits_{n=0}^{\infty}(1+q^{n+\frac12}z)(1+q^{n+\frac12}z^{-1})(1-q^{n+1})\\
\vartheta_4(\tau,\nu)&\ =\prod\limits_{n=0}^{\infty}(1-q^{n+\frac12}z)(1-q^{n+\frac12}z^{-1})(1-q^{n+1})\\
\eta(\tau)&\ =q^{\frac{1}{24}}\prod_{n=0}^{\infty}(1-q^{n+1})\ .
\end{split}
\end{equation*}

\subsection[$c=3$ characters]{$\boldsymbol{c=3}$ characters}

We discuss here the characters of the unitary fully supersymmetric
irreducible representations of the $N=2$ superconformal algebra at
$c=3$. The Verma modules of the $N=2$ superconformal algebra contain several
singular submodules,\footnote{In general there are also subsingular
submodules, but they do not show up for unitary
representations~\cite{Klemm:2003vn}.} which have to be taken into
account. The structure of the singular submodules can be read off from
the embedding diagrams of the representations (for further details we
refer to~\cite{Kiritsis:1986rv,Eholzer:1996zi}); we will follow the
classification of~\cite{Klemm:2003vn}. Let us explain the procedure at
the example of the characters for the representations of type
$I^{\pm}$ in the notations of the aforementioned paper; the labels
satisfy $\frac{h}{Q}\in \Z+\frac12$, with positive $h$ and
$Q\not\in\mathbb{Z}$. In this case we have only one charged singular
vector. The singular vectors at level
$\frac{h}{|Q|}=n+\frac12$ can be recognised to be\footnote{One can,
for instance, follow the spectral flow of Neveu-Schwarz null vectors
starting from the (anti)chiral primaries.}
\begin{align}
\begin{array}{ll}
G^-_{\frac12}G^-_{\frac32}\dots G^{-}_{\frac{h}{|Q|}-1}G^{+}_{-\frac{h}{|Q|}}G^{+}_{-\frac{h}{|Q|}+1}\dots G^{+}_{-\frac32} G^{+}_{-\frac12}|n,Q\ket\quad \text{for}\ Q>0\\
G^+_{\frac12}G^+_{\frac32}\dots
G^{+}_{\frac{h}{|Q|}-1}G^{-}_{-\frac{h}{|Q|}}G^{-}_{-\frac{h}{|Q|}+1}\dots
G^{-}_{-\frac32} G^{-}_{-\frac12}|n,Q\ket\quad \text{for}\ Q<0
\end{array}\ ,
\end{align}
and they have relative charge $+1$ and $-1$, respectively. Here,
$G^{\pm}_{r}$ denote the modes of the supercurrents of the $N=2$
superconformal algebra. In the character of the irreducible
representation we have to subtract the
contribution of the submodule associated to them. The result is
\begin{equation}
\chi^{I^{\pm}}_{n,Q}(q,z)=q^{(n+\frac12)|Q|-\frac18}z^Q\left[\prod_{m=0}^{\infty}\frac{(1+q^{m+\frac12}z)(1+q^{m+\frac12}z^{-1})}{(1-q^{m+1})^2}\right]\left(1-\frac{q^{n+\frac12}z^{\text{sgn}Q}}{1+q^{n+\frac12}z^{\text{sgn}Q}}\right)\ .
\end{equation}
The other cases are analogous, and we can write:
\begin{subequations}\label{c=3-characters}
\begin{itemize}
\item \textbf{Vacuum:} ($Q=h=0$) 
\begin{flalign}
&\chi^{\text{vac}}_{0,0}(q,z)= \frac{\vartheta_{3}(\tau,\nu)}{\eta^{3}(\tau)} 
\left(1-\frac{q^{\frac12}z}{1+q^{\frac12}z}-\frac{q^{\frac12}z^{-1}}{1+q^{\frac12}z^{-1}}\right)&
\end{flalign}
\item \textbf{Type ${\boldsymbol{0}}$:} ($Q=0\, ,\ h\in\R\setminus\{0\}$)
\begin{flalign}
&\chi^{0}_{h,0}(q,z)=q^{h}
\frac{\vartheta_{3}(\tau,\nu)}{\eta^{3}(\tau)}  &
\end{flalign}
\item \textbf{Type ${\boldsymbol{I^{\pm}}}$:} ($0<|Q|<1\, ,\
h=|Q|(n+\frac12)\, ,\  n\in \Z_{\geq 0}$)
\begin{flalign}
&\chi^{I^{\pm}}_{|Q|(n+\frac12),Q}(q,z)=q^{(n+\frac12)|Q|}z^Q
\frac{\vartheta_{3}(\tau,\nu)}{\eta^{3}(\tau)}  
\left(1-\frac{q^{n+\frac12}z^{\text{sgn}Q}}{1+q^{n+\frac12}z^{\text{sgn}Q}}\right)&
\end{flalign}
\item \textbf{Type ${\boldsymbol{II}}^{\pm}$:} ($Q=\pm1\, ,\
h\in\R_{\geq0}$)
\begin{flalign}
&\chi^{II^{\pm}}_{h,Q}(q,z)=q^{h}z^Q
\frac{\vartheta_{3}(\tau,\nu)}{\eta^{3}(\tau)} \left(1-q^{|Q|}\right)&
\end{flalign}
\item \textbf{Type ${\boldsymbol{III^{\pm}}}$:} ($Q=\pm1\, ,\
h\in\Z +\frac12$)
\begin{flalign}
&\chi^{III^{\pm}}_{h,Q}(q,z)=q^{h}z^Q
\frac{\vartheta_{3}(\tau,\nu)}{\eta^{3}(\tau)}
\left(1-q-
\frac{q^{h}z^{\text{sgn}(Q)}}{1+q^{h}z^{\text{sgn}(Q)}}+
\frac{q^{h+2}z^{\text{sgn}(Q)}}{1+q^{h+1}z^{\text{sgn}(Q)}}
\right)&
\end{flalign}
\end{itemize}
\end{subequations}

Ramond characters can be obtained from the Neveu-Schwarz characters by
spectral flow (see e.g.\ \cite{Lerche:1989uy}). We
give an example: let us denote spectral flowed operators and sectors
by an upper label $\eta$, which indicates the amount of spectral flow
units to use. Under a flow of $\eta=\pm 1/2$, primary vectors of the
Neveu-Schwarz sector become Ramond
primaries, and the same happens for Neveu-Schwarz
singular vectors, which flow to Ramond singular vectors. The Ramond
characters can then be computed using the formula
\begin{equation}
\chi_{h^{\eta},Q^{\eta}}^{\phantom {\text{NS}}}(q,z)=\Tr_{\mathcal{H}_{h^{\eta},Q^{\eta}}}q^{L_0-\frac{c}{24}}z^{J_0}=\Tr_{\mathcal{H}_{h,Q}}q^{L^{\eta}_0-\frac{c}{24}}z^{J^{\eta}_0}\ ,
\end{equation}
with the spectral flowed operators
\begin{align}
 L_n^{\eta}&=L_n-\eta J_n+\frac{c}{6}\eta^2\delta_{n,0} &
 J_n^{\eta}&=J_n-\frac{c}{3}\eta\delta_{n,0}\ .
\end{align}
For $c=3$ and $\eta=\frac12$, $L^{1/2}_{0}=L_0-\frac12J_0+\frac18$ and $J^{1/2}_0=J_0-\frac12$, we have
\begin{equation}
\chi_{h^{1/2},Q^{1/2}}^{\phantom {\text{NS}}}(q,z)=q^{\frac18}z^{-\frac12}\chi_{h,Q}(q,q^{-\frac12}z)\ .
\end{equation}
Starting e.g.\ from the type $I$ characters in the Neveu-Schwarz
sector we find the characters
\begin{align}\label{R0-character}
\chi^{\text{R}^0}_{\frac18,Q}(q,z)&=\frac{z^{Q}}{z^{1/2}-z^{-1/2}}
\frac{\vartheta_{2} (\tau,\nu)}{\eta^{3} (\tau)} \ ,\
-\frac{1}{2}<Q<\frac{1}{2} \\
\chi^{\text{R}}_{\frac18+n|Q|,Q}(q,z)&=
\frac{q^{n|Q|}z^{Q}}{1+q^{n}z^{\text{sgn}(Q)}}
\frac{\vartheta_{2} (\tau,\nu)}{\eta^{3} (\tau)}  \ ,\ 0<|Q|<1 \ ,\
n\geq 1\ ,
\end{align}
where in the first character the lowest lying state is a Ramond ground
state, whereas in the second character there are two lowest lying
states of charges $Q\pm \frac{1}{2}$.

\subsection{Minimal model characters and partition function}
\label{app:minimalmodel} 
Unitary irreducible representations for the bosonic subalgebra of the
$N=2$ superconformal algebra at central charge $c=3\frac{k}{k+2}$ are
labelled by three integers $(l,m,s)$ with $0\leq l\leq k$, $m\equiv
m+2k+4$, $s\equiv s+4$, and $l+m+s$ even. Not all triples label
independent representations, and they are identified according to
\begin{equation}\label{fieldidentification}
(l,m,s) \sim (k-l,m+k+2,s+2) \ .
\end{equation}
Representations of the full superconformal algebra are then
obtained by combining representations labelled by $(l,m,s)$ and $(l,m,s+2)$.

Explicit expressions for the characters of the $N=2$ superconformal
algebra can be found e.g. in~\cite{Ravanini:1987yg}. In the 
Neveu-Schwarz sector for $|m|\leq l$ they read
\begin{align}
\chi^{\text{NS}}_{l,m}(q,z):=&\left( \chi_{(l,m,0)}+ \chi_{(l,m,2)}\right)(q,z)\nonumber\\
= &\ q^{\frac{(l+1)^{2}-m^2}{4(k+2)}-\frac18}\,z^{-\frac{m}{k+2}}\left[\prod_{n=0}^{\infty}\frac{(1+q^{n+\frac12}z)(1+q^{n+\frac12}z^{-1})}{(1-q^{n+1})^2}\right]\times \Gamma^{(k)}_{lm}(\tau,\nu)\ ,
\label{minmod-NS-character}
\end{align}
and in the Ramond sector (for $|m|\leq l+1$)
\begin{align}
\chi^{R}_{l,m}(q,z):=&\left(\chi_{(l,m,1)}+\chi_{(l,m,-1)}\right)(q,z)\nonumber\\
= \ &q^{\frac{(l+1)^{2}-m^2}{4(k+2)}}\,z^{-\frac{m}{k+2}}(z^{\frac12}+z^{-\frac12})\left[\prod_{n=0}^{\infty}\frac{(1+q^{n+1}z)(1+q^{n+1}z^{-1})}{(1-q^{n+1})^2}\right]\times \Gamma^{(k)}_{lm}(\tau,\nu)\ ,
\label{minmod-R-characters}
\end{align}
where the structure of the singular vectors is summarised in $\Gamma^{(k)}_{lm}$,
\begin{align}
 \Gamma^{(k)}_{lm}(\tau,\nu) = &\sum_{p=0}^{\infty}q^{(k+2)p^2+(l+1)p}\left(1-\frac{q^{(k+2)p+\frac{l+m+1}{2}}z}{1+q^{(k+2)p+\frac{l+m+1}{2}}z}-\frac{q^{(k+2)p+\frac{l-m+1}{2}}z^{-1}}{1+q^{(k+2)p+\frac{l-m+1}{2}}z^{-1}}\right)\nonumber\\
  - & \sum_{p=1}^{\infty}q^{(k+2)p^2-(l+1)p}\left(1-\frac{q^{(k+2)p-\frac{l+m+1}{2}}z^{-1}}{1+q^{(k+2)p-\frac{l+m+1}{2}}z^{-1}}-\frac{q^{(k+2)p-\frac{l-m+1}{2}}z}{1+q^{(k+2)p-\frac{l-m+1}{2}}z}\right)\ .
\label{def-Gamma}
\end{align}
The Neveu-Schwarz part of the minimal model partition function is
given by
\begin{equation}\label{Z-MM-A}
\begin{split}
Z^{\text{NS}}_{k}(\tau,\nu)=&\sum_{l=0}^{k}\sum_{\substack{m=-l\\ l+m\ \text{even}}}^l\chi_{(l,m,0)} (q,z)\bar\chi_{(l,m,0)} (\bar{q},\bar{z}) +\chi_{(l,m,2)} (q,z)\bar\chi_{(l,m,2)} (\bar{q},\bar{z})\\
=&\frac12\sum_{l=0}^{k}\sum_{\substack{m=-l\\ l+m\
\text{even}}}^l\left(\chi^{\text{NS}}_{l,m}
(q,z)\bar\chi^{\text{NS}}_{l,m}(\bar{q},\bar{z})+\chi^{\text{NS}}_{l,m}
(q,-z)\bar\chi^{\text{NS}}_{l,m}(\bar{q},-\bar{z})\right)\ .
\end{split}
\end{equation}
It can be seen as a (GSO-like) projection of the trace over the full
supersymmetric Neveu-Schwarz Hilbert space
$\cH^{\text{NS}}_{k}=\oplus_{|m|\leq l\leq k}\cH^{\text{NS}}_{l,m}\otimes
\cH^{\text{NS}}_{l,m}$,
\begin{equation}\label{Z-MM-SUSY}
\begin{split} 
\mathcal{P}_{k}^{\text{NS}}(\tau,\nu):&=\sum_{l=0}^{k}\sum_{\substack{m=-l\\ l+m\ \text{even}}}^l\left(\chi_{(l,m,0)} (q,z)+\chi_{(l,m,2) (q,z)}\right)\left(\bar\chi_{(l,m,0)} (\bar{q},\bar{z})+\bar\chi_{(l,m,2) (\bar{q},\bar{z})}\right)\\
&=\left|\frac{\vartheta_3(\tau,\nu)}{\eta^3(\tau)}\right|^2\sum_{l=0}^{k}\sum_{\substack{m=-l\\
l+m\ \text{even}}}^l\left|q^{\frac{(l+1)^{2}-m^2}{4(k+2)}}
\Gamma^{(k)}_{lm}(\tau,\nu)\right|^2\, (z\bar{z})^{-\frac{m}{k+2}}\ .
\end{split}
\end{equation}

\subsection{Limit of minimal model characters}\label{app:limit-char}

In the limit $k\to\infty$ in the expression~\eqref{def-Gamma} for
$\Gamma^{(k)}_{lm}$ in each sum only the first summand can contribute,
\begin{align}\label{limit-Gamma}
 \Gamma^{(k)}_{lm}(\tau,\nu) \approx &\left( 1-\frac{q^{\frac{l+m+1}{2}}z}{1+q^{\frac{l+m+1}{2}}z}-\frac{q^{\frac{l-m+1}{2}}z^{-1}}{1+q^{\frac{l-m+1}{2}}z^{-1}}\right)\nonumber\\
 & - q^{k-l+1}\left(1-\frac{q^{\frac{2k-l-m+3}{2}}z^{-1}}{1+q^{\frac{2k-l-m+3}{2}}z^{-1}}-\frac{q^{\frac{2k-l+m+3}{2}}z}{1+q^{\frac{2k-l+m+3}{2}}z}\right)
 \ , 
\end{align}
and the precise behaviour of the character depends on the details
of how $l$ and $m$ behave in the limit.

For our analysis we need to consider the following cases in the Neveu-Schwarz sector:
\begin{enumerate}
\item $l=m=0$: The limit character is simply the $N=2$ vacuum
character,
\begin{equation}
\lim_{k\to\infty} \chi^{\text{NS}}_{0,0} = \chi^{\text{vac}}_{0,0}\ .
\end{equation}
\item $l+m=2n$ finite, $m/ (k+2)\to -Q$, $0<Q<1$: Only one singular
vector survives and we find
\begin{equation}
\lim_{k\to \infty} \chi^{\text{NS}}_{|m|+2n,m}
= \chi_{Q (n+\frac{1}{2}),Q}^{I^{+}} \ .
\end{equation}
\item $l-m=2n$ finite, $m/ (k+2)\to -Q$, $-1<Q<0$: Only one singular
vector survives and we find
\begin{equation}
\lim_{k\to \infty} \chi^{\text{NS}}_{m+2n,m}
= \chi_{|Q| (n+\frac{1}{2}),Q}^{I^{-}} \ .
\end{equation}
\item $l+m=2n$ finite, $l=k$: The first summand in~\eqref{limit-Gamma}
gives one positively charged singular vector, the second produces one uncharged one and adds one positively charged singular submodule. We find
\begin{equation}\label{limit-III-minus}
\lim_{k\to\infty} \chi^{\text{NS}}_{k,-k+2n} 
= \chi_{n+\frac{1}{2},1}^{III^{+}} \ .
\end{equation}
\item $l-m=2n$ finite, $l=k$: Analogously to the previous case we obtain
\begin{equation}\label{limit-III-plus}
\lim_{k\to\infty} \chi^{\text{NS}}_{k,k-2n} 
= \chi_{n+\frac{1}{2},-1}^{III^{-}} \ .
\end{equation}
\end{enumerate}
There are several other cases, depending on the behaviour of $l\pm m$
for large $k$; in these other situations the limiting character
decomposes into a sum of $N=2$ characters. We illustrate this in the
example of fixed labels $l,m$: in this instance the
conformal weights and $U(1)$ charge of all the primary fields approach
zero, the second line of equation~\eqref{limit-Gamma} gets suppressed,
but the first line stays finite. The character then takes the
form
\begin{equation}
\lim_{k\to\infty}\chi^{\text{NS}}_{l,m} (q,z) =\frac{\vartheta_3(\tau,\nu)}{\eta^3(\tau)}\left( 1-\frac{q^{\frac{l+m+1}{2}}z}{1+q^{\frac{l+m+1}{2}}z}-\frac{q^{\frac{l-m+1}{2}}z^{-1}}{1+q^{\frac{l-m+1}{2}}z^{-1}}\right)\ .
\end{equation}
Noticing the relation
\begin{equation}
\frac{\vartheta_3(\tau,\nu)}{\eta^3(\tau)}
\left(\frac{q^{n+\frac{1}{2}}z^{\pm 1}}{1+q^{n+\frac{1}{2}}z^{\pm 1}}
- \frac{q^{n+\frac{3}{2}}z^{\pm 1}}{1+q^{n+\frac{3}{2}}z^{\pm 1}}
\right) = \chi^{III^{\pm}}_{n+\frac{1}{2},\pm 1} (q,z) \ ,
\end{equation}
it is easy to show that
\begin{equation}
\lim_{k\to\infty}\chi^{\text{NS}}_{l,m} =
\chi^{\text{vac}}_{0,0}+\sum_{j=0}^{\frac{l+m}{2}-1}\chi^{III^+}_{\frac{l+m}{2}-(\frac12+j),1}+\sum_{j=0}^{\frac{l-m}{2}-1}\chi^{III^-}_{\frac{l-m}{2}-(\frac12+j),-1}\ .
\end{equation}
Following similar lines it is possible to show that this kind of
decomposition is common to all the cases we have not listed
explicitly.

\section{Asymptotics of Wigner 3j-symbols}\label{app:3j}

\begin{figure}[h!]
  \centering
    \includegraphics{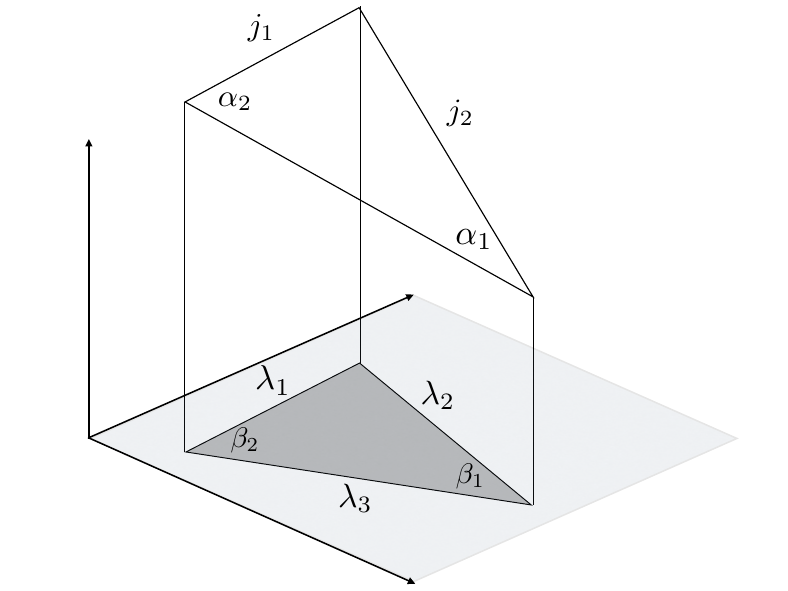}
\caption{\label{fig:PR-angles}Ponzano-Regge angles defined in
equations~(\ref{PR-angles}) and~(\ref{def-alphas}): the $\alpha_i$ are
the internal angles of the triangle formed by the $j_i$ labels; the
$\beta_i$ are the internal angles of the triangle projected on the
xy-plane (where the $\mu_{i}$ measure the z-components of the angular momenta); $\gamma_i$ (not present here) is the angle between the outer normals to the faces adjacent to the edge $j_i$.}
\end{figure}

We are interested in the region of the parameter space of 3j-symbols
in which the angular momentum labels $j_{i}$ scale like $j_i \propto \sqrt{k}$ and the
magnetic labels $\mu_i$ stay finite in the limit of large $k$. In this
range we are deeply inside the classically allowed region (see
e.g.\ the appendix~A of~\cite{Fredenhagen:2012rb} for more details),
and we can use the approximation methods derived
in~\cite{Reinsch:1999}. In particular we find there \cite[eq.\ (3.23)]{Reinsch:1999}
\begin{equation}\label{3j-deep}
\begin{pmatrix}
j_1 & j_2 & j_3 \\
\mu_1 & \mu_2 & \mu_3
\end{pmatrix}
\simeq
2 I_{j_1 \, \mu_1 \, j_2 \, \mu_2 \, j_3 \, \mu_3} \,
(-1)^{j_1-j_2-\mu_3}\sqrt{\frac{j_3}{2j_3+1}}\frac{\cos\left[\chi + \frac{\pi}{4} - \pi(j_3+1) \right]
}{(4\pi A (\lambda_{1},\lambda_{2},\lambda_{2}))^{1/2}}\ ,
\end{equation}
where $\chi$ is defined as
\begin{equation}\label{chi-def}
 \chi =(j_1+\tfrac{1}{2})\gamma_1+(j_2+\tfrac{1}{2})
 \gamma_2+(j_3+\tfrac{1}{2})
 \gamma_3+\mu_2 \beta_1-\mu_1 \beta_2\ .
\end{equation}
We use the Ponzano-Regge angles $\gamma_{1,2,3},\beta_{1,2}$
(see~\cite{Ponzano:1968} and figure~\ref{fig:PR-angles}) which through
their cosines read
\begin{subequations}\label{PR-angles}
\begin{align}
\cos\gamma_1 &= {\frac{\mu_3(j_1^2 + j_2^2 - j_3^2) - \mu_2(j_1^2 + j_3^2 -j_2^2)}
{4A (j_{1},j_{2},j_{3}) \lambda_1}} & 
\cos\beta_1 &= {\frac{\lambda_3^2 + \lambda_2^2 - \lambda_1^2}
{2 \lambda_2 \lambda_3}}\\[5pt]
\cos\gamma_2 &= {\frac{\mu_1(j_3^2 + j_2^2 - j_1^2) - \mu_3(j_2^2 + j_1^2 -j_3^2)}
{4A (j_{1},j_{2},j_{3}) \lambda_2}} & 
\cos\beta_2 &= {\frac{\lambda_1^2 + \lambda_3^2 - \lambda_2^2}
{2 \lambda_1 \lambda_3}}\\[5pt]
\cos\gamma_3 &= {\frac{\mu_2(j_1^2 + j_3^2 - j_2^2) - \mu_1(j_3^2 + j_2^2 -j_1^2)}
{4A (j_{1},j_{2},j_{3}) \lambda_3}} \ .& &  
\end{align}
\end{subequations}
Here,
\begin{align}
\lambda_i &= \sqrt{j_i^2-\mu_i^2} \qquad i = 1,2,3 \ ,\\
\intertext{and}
A (x_{1},x_{2},x_{3}) &=\frac{1}{4}
\sqrt{(x_3 + x_1 + x_2) (-x_3 + x_1 + x_2) (x_3 - x_1 + x_2) (x_3 + x_1 - x_2)}
\label{area}
\end{align}
is the area of the triangle with side lengths $x_{i}$.

The quantity $I_{j_1 \, \mu_1 \, j_2 \, \mu_2 \, j_3 \, \mu_{3}}$ appearing in
equation~\eqref{3j-deep} is defined as
\begin{multline}
I_{j_1 \, \mu_1 \, j_2 \, \mu_2 \, j_3 \, \mu_3} =
\sqrt{ \frac{(j_3+1/2)(j_3+j_1+j_2)}{j_3(j_3+j_1+j_2+1)} } \\
\times\ 
\frac{f(j_1+\mu_1)\, f(j_1-\mu_1)\, f(j_2+\mu_2)\, f(j_2-\mu_2)\,
f(j_3+\mu_3)\,f(j_3-\mu_3)}
{f(j_1+j_2+j_3)\, f(j_1+j_2-j_3)\, f(j_1-j_2+j_3)\, f(-j_1+j_2+j_3)} \  ,
\end{multline}
where $f(n)$ is the square root of the ratio of $n!$ to the Stirling
approximation of $n!$, and has the following large $n$ behaviour
\begin{equation}
f(n) = \sqrt{\frac{n!}{\sqrt{2 \pi n}\,n^n e^{-n}}} =1+\frac{1}{24 n}+\mathcal{O} \left(\frac{1}{n^2}\right)\ .
\end{equation}

We now consider the situation where the labels $j_{i}$ are
proportional to $\sqrt{k}$ for large $k$ while keeping $\mu_i$
finite. In this regime we have
\begin{equation}
I = 1+\mathcal O\left(k^{-1/2}\right)\ ,
\end{equation}
and the angles behave as follows:
\begin{subequations}
\begin{align}
\cos\gamma_{1,2,3} & = f_{1,2,3}+\mathcal{O}(k^{-3/2})\\
\cos\beta_1 & =\frac{-j_1^2+j_2^2+j_3^2}{2j_2j_3}+\mathcal{ O}(k^{-1})
& \cos\beta_2&=\frac{j_1^2-j_2^2+j_3^2}{2j_1j_3}+\mathcal {O}
(k^{-1})\ ,
\end{align}
\end{subequations}
where we used the definitions
\begin{subequations}\label{def-fs}
\begin{alignat}{3}
 f_1&=\frac{\mu_3(j_1^2+j_2^2-j_3^2)-\mu_2(j_1^2-j_2^2+j_3^2)}{4A (j_{1},j_{2},j_{3}) j_1} &&\propto k^{-\frac{1}{2}}\\
 f_2&=\frac{\mu_1(-j_1^2+j_2^2+j_3^2)-\mu_3(j_1^2+j_2^2-j_3^2)}{4A (j_{1},j_{2},j_{3}) j_2}&&\propto k^{-\frac{1}{2}}\\
 f_3&=\frac{\mu_2(j_1^2-j_2^2+j_3^2)-\mu_1(-j_1^2+j_2^2+j_3^2)}{4A (j_{1},j_{2},j_{3}) j_3}&&\propto k^{-\frac{1}{2}}\ .
\end{alignat}
\end{subequations}
Inverting~\eqref{PR-angles} and expanding in $k$ we get
$\gamma_{1,2,3}=\frac{\pi}{2}-\frac{f_{1,2,3}}{k^{1/2}}+\mathcal{O}(k^{-3/2})$,
so that $\chi$ of \mbox{eq.\ \eqref{chi-def}} becomes
\begin{align}
 \chi = \frac{\pi}{2}(j_1+j_2+j_3) 
+ \bigg[\frac34\pi&-(j_1f_1+j_2f_2+j_3f_3)-\mu_1\cos^{-1}\frac{j_1^2-j_2^2+j_3^2}{2j_1j_3}\nonumber\\
&+ \mu_2\cos^{-1}\frac{-j_1^2+j_2^2+j_3^2}{2j_2j_3}\bigg]+\mathcal{O}(k^{-1/2})\ .
\end{align}
Since $\sum j_i f_i=0$, we have
\begin{equation}
\begin{split}
&\cos\left[\chi + \frac{\pi}{4} - \pi(j_3+1)
\right]=\cos\left[(j_1+j_2-j_3)\frac{\pi}{2}+\mu_2\alpha_1-\mu_1\alpha_2\right]
\end{split}
\end{equation}
with (see figure~\ref{fig:PR-angles})
\begin{align}\label{def-alphas}
 \alpha_1&:=\arccos \frac{-j_1^2+j_2^2+j_3^2}{2j_2 j_3} &
 \alpha_2&:=\arccos \frac{j_1^2-j_2^2+j_3^2}{2j_1 j_3}\ .
\end{align}
The remaining factor behaves as $\sqrt{\frac{j_3}{2j_3+1}}=\frac{1}{\sqrt2}\left(1+\mathcal O (k^{-1/2})\right)$.\\
Collecting all the pieces we get
\begin{equation}\label{3j-asympt-app}
\begin{pmatrix}
j_1 & j_2 & j_3 \\
\mu_1 & \mu_2 & \mu_3
\end{pmatrix} 
=\frac{(-1)^{j_1-j_2-\mu_3}}{\sqrt{2\pi A
(j_{1},j_{2},j_{3})}}\cos\left[(j_1+j_2-j_3)\frac{\pi}{2}+\mu_2\alpha_1-\mu_1\alpha_2
\right]\left(1+\mathcal{O}(k^{-1/2}) \right)\ .
\end{equation}

\section{Free field three-point function}
\label{sec:app-free-three}

In the supersymmetric free field theory of two bosons and two fermions
on the plane the three-point function of Neveu-Schwarz (super-)primary
fields $\Phi^{\text{free}}_{\mathbf{p}}$ is very simple
(see \mbox{eq.\ \eqref{freefield}}). In this appendix we compute the three-point
function in a radial basis $\Phi^{\text{free}}_{p,m}$, which is needed
for the comparison to the limit of minimal models.

\begin{figure}
\begin{center}
\includegraphics{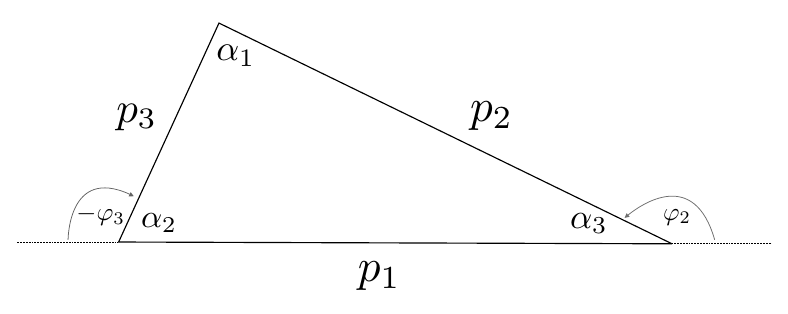}
\end{center}
\caption{\label{fig:free-angles}The triangle spanned by $p_{1}$,
$p_{2}e^{i\varphi_{2}}$ and $p_{3}e^{i\varphi_{3}}$.}
\end{figure}

The fields $\Phi^{\text{free}}_{p,m}$ are defined as 
\begin{equation}
\Phi^{\text{free}}_{p,m} = \sqrt{\frac{p}{2\pi}}\int d\varphi\ 
\Phi^{\text{free}}_{pe^{i\varphi}}\, e^{im\varphi} \ .
\end{equation}   
Their three-point function can therefore be expressed as
\begin{align}
&\langle \Phi^{\text{free}}_{p_{1},m_{1}} (z_{1},\bar{z}_{1})
\Phi^{\text{free}}_{p_{2},m_{2}} (z_{2},\bar{z}_{2})
\Phi^{\text{free}}_{p_{3},m_{3}} (z_{3},\bar{z}_{3})\rangle\nonumber\\
&\  = \sqrt{\frac{p_{1}p_{2}p_{3}}{(2\pi)^{3}}} \int d\varphi_{1}
d\varphi_{2} d\varphi_{3}\
e^{im_{1}\varphi_{1}+im_{2}\varphi_{2}+im_{3}\varphi_{3}}
 \langle \Phi^{\text{free}}_{p_{1}e^{i\varphi_{1}}} (z_{1},\bar{z}_{1})
\Phi^{\text{free}}_{p_{2}e^{i\varphi_{2}}} (z_{2},\bar{z}_{2})
\Phi^{\text{free}}_{p_{3}e^{i\varphi_{3}}} (z_{3},\bar{z}_{3})\rangle\nonumber\\
&\  =  \sqrt{\frac{p_{1}p_{2}p_{3}}{(2\pi)^{3}}} |z_{12}|^{2 (h_{3}-h_{1}-h_{2})} |z_{23}|^{2 (h_{1}-h_{2}-h_{3})} 
 |z_{13}|^{2 (h_{2}-h_{1}-h_{3})}\nonumber\\
&\qquad \times \int d\varphi_{1}
d\varphi_{2} d\varphi_{3}\
e^{im_{1}\varphi_{1}+im_{2}\varphi_{2}+im_{3}\varphi_{3}}\, 
\delta^{(2)}
(p_{1}e^{i\varphi_{1}}+p_{2}e^{i\varphi_{2}}+p_{3}e^{i\varphi_{3}}) \\
&\ = \sqrt{\frac{p_{1}p_{2}p_{3}}{2\pi}} |z_{12}|^{2 (h_{3}-h_{1}-h_{2})} |z_{23}|^{2 (h_{1}-h_{2}-h_{3})} 
 |z_{13}|^{2 (h_{2}-h_{1}-h_{3})}\nonumber\\
&\qquad \times \delta_{m_{1}+m_{2}+m_{3}}\int
d\varphi_{2} d\varphi_{3}\
e^{im_{2}\varphi_{2}+im_{3}\varphi_{3}}\,
\delta^{(2)}(p_{1}+p_{2}e^{i\varphi_{2}}+p_{3}e^{i\varphi_{3}}) \ .
\end{align}
We now have to evaluate the remaining integral over the angles
$\varphi_{2}$ and $\varphi_{3}$. Due to the delta-distribution it only gets
contributions if the two-dimensional vectors corresponding to the
complex momenta $p_{1}$, $p_{2}e^{i\varphi_{2}}$ and
$p_{3}e^{i\varphi_{3}}$ form a triangle. In particular it is zero unless the inequalities
\begin{equation}
|p_{2}-p_{3}| \leq p_{1} \leq p_{2}+p_{3}
\end{equation}
are satisfied. The triangle condition arising from the
delta-distribution can be formulated by the
equations
\begin{alignat}{2}
q_{1} &:= p_{1} + p_{2}\cos \varphi_{2} + p_{3} \cos \varphi_{3} && = 0\\
q_{2} &:= p_{2}\sin \varphi_{2} + p_{3}\sin \varphi_{3} && = 0  \ .
\end{alignat}
The angles $\varphi_{i}$ take values in the interval $[-\pi,\pi]$. For
any solution $(\varphi_{2},\varphi_{3})$ there is another solution
$(-\varphi_{2},-\varphi_{3})$ that corresponds to the triangle reflected
at the side $p_{1}$. For $\varphi_{2}>0$ we have $\varphi_{3}<0$ and
the relation to the angles of the triangle is given by (see figure~\ref{fig:free-angles})
\begin{align}
\varphi_{2} &= \alpha_{1}+\alpha_{2}  &
\varphi_{3} &= \alpha_{2} -\pi \ .
\end{align}
Evaluating the integral therefore reduces to plugging in the values
for $\varphi_{2}$ and $\varphi_{3}$ for the two solutions, and
dividing this by the Jacobian determinant
\begin{equation}
\left| \det \left( \frac{\partial q_{i}}{\partial \varphi_{j}} \right)_{i,j}\right| = 
p_{2} p_{3} |\sin (\varphi_{2}-\varphi_{3})| = 2 A (p_{1},p_{2},p_{3})
\ ,
\end{equation}
where $A (p_{1},p_{2},p_{3})$ is the area of the triangle (see eq.\
\eqref{area}).
We find in total
\begin{align}
&\langle \Phi^{\text{free}}_{p_{1},m_{1}} (z_{1},\bar{z}_{1})
\Phi^{\text{free}}_{p_{2},m_{2}} (z_{2},\bar{z}_{2})
\Phi^{\text{free}}_{p_{3},m_{3}} (z_{3},\bar{z}_{3})\rangle\nonumber\\
&\quad = \sqrt{\frac{p_{1}p_{2}p_{3}}{2\pi}} |z_{12}|^{2 (h_{3}-h_{1}-h_{2})} |z_{23}|^{2 (h_{1}-h_{2}-h_{3})} 
 |z_{13}|^{2 (h_{2}-h_{1}-h_{3})}\nonumber\\
&\qquad \times \delta_{m_{1}+m_{2}+m_{3}}
\frac{\cos (m_{2}\alpha_{1}-m_{1}\alpha_{2} +\pi (m_{1}+m_{2}))}{A
(p_{1},p_{2},p_{3})}\ .
\end{align}



\begin{thebibliography}{10}

\bibitem{Runkel:2001ng}
I.~Runkel, G.~Watts, {\em {A Nonrational CFT with $c = 1$ as a limit of minimal
  models}\/}, JHEP {\bf 0109} (2001) 006,
  \href{http://xxx.lanl.gov/abs/hep-th/0107118}{{\ttfamily
  arXiv:hep-th/0107118}}

\bibitem{Graham:2001tg}
K.~Graham, I.~Runkel, G.~Watts, {\em {Minimal model boundary flows and $c=1$
  {CFT}}\/}, Nucl.Phys. {\bf B608} (2001) 527,
  \href{http://xxx.lanl.gov/abs/hep-th/0101187}{{\ttfamily
  arXiv:hep-th/0101187}}

\bibitem{Roggenkamp:2003qp}
D.~Roggenkamp, K.~Wendland, {\em {Limits and degenerations of unitary conformal
  field theories}\/}, Commun.Math.Phys. {\bf 251} (2004) 589,
  \href{http://xxx.lanl.gov/abs/hep-th/0308143}{{\ttfamily
  arXiv:hep-th/0308143}}

\bibitem{Fredenhagen:2004cj}
S.~Fredenhagen, V.~Schomerus, {\em {Boundary Liouville theory at $c = 1$}\/},
  JHEP {\bf 0505} (2005) 025,
  \href{http://xxx.lanl.gov/abs/hep-th/0409256}{{\ttfamily
  arXiv:hep-th/0409256}}

\bibitem{Fredenhagen:2007tk}
S.~Fredenhagen, D.~Wellig, {\em {A common limit of super Liouville theory and
  minimal models}\/}, JHEP {\bf 0709} (2007) 098,
  \href{http://xxx.lanl.gov/abs/0706.1650}{{\ttfamily arXiv:0706.1650}}

\bibitem{Roggenkamp:2008jm}
D.~Roggenkamp, K.~Wendland, {\em {Decoding the geometry of conformal field
  theories}\/}, Bulg.J.Phys. {\bf 35} (2008) 139,
  \href{http://xxx.lanl.gov/abs/0803.0657}{{\ttfamily arXiv:0803.0657}}

\bibitem{Fredenhagen:2010zh}
S.~Fredenhagen, {\em {Boundary conditions in Toda theories and minimal
  models}\/}, JHEP {\bf 1102} (2011) 052,
  \href{http://xxx.lanl.gov/abs/1012.0485}{{\ttfamily arXiv:1012.0485}}

\bibitem{Fredenhagen:2012rb}
S.~Fredenhagen, C.~Restuccia, R.~Sun, {\em {The limit of N=(2,2) superconformal
  minimal models}\/}, JHEP {\bf 1210} (2012) 141,
  \href{http://xxx.lanl.gov/abs/1204.0446}{{\ttfamily arXiv:1204.0446}}

\bibitem{Zamolodchikov:1986gt}
A.~Zamolodchikov, {\em {Irreversibility of the Flux of the Renormalization
  Group in a 2D Field Theory}\/}, JETP Lett. {\bf 43} (1986) 730

\bibitem{Kutasov:1988xb}
D.~Kutasov, {\em {Geometry on the space of conformal field theories and contact
  terms}\/}, Phys.Lett. {\bf B220} (1989) 153

\bibitem{Douglas:2010ic}
M.~R. Douglas, {\em {Spaces of Quantum Field Theories}\/}  (2010),
  \href{http://xxx.lanl.gov/abs/1005.2779}{{\ttfamily arXiv:1005.2779}}

\bibitem{Maldacena:2001ky}
J.~M. Maldacena, G.~W. Moore, N.~Seiberg, {\em {Geometrical interpretation of
  D-branes in gauged WZW models}\/}, JHEP {\bf 0107} (2001) 046,
  \href{http://xxx.lanl.gov/abs/hep-th/0105038}{{\ttfamily
  arXiv:hep-th/0105038}}

\bibitem{Andrews:book}
G.~E. Andrews, R.~Askey, R.~Roy, {\em Special Functions\/}, number~71 in
  Encyclopedia of Mathematics and its Applications, Cambridge University Press,
  Cambridge (1999)

\bibitem{BlumenhagenLuestTheisen}
R.~Blumenhagen, D.~L{\"u}st, S.~Theisen, {\em Basic concepts of string
theory}, Springer, Heidelberg (2013)

\bibitem{Mussardo:1988av}
G.~Mussardo, G.~Sotkov, M.~Stanishkov, {\em N=2 superconformal minimal
  models\/}, Int. J. Mod. Phys. {\bf A4} (1989) 1135

\bibitem{Zamolodchikov:1986bd}
A.~B. Zamolodchikov, V.~A. Fateev, {\em Operator algebra and correlation
  functions in the two-dimensional Wess-Zumino SU(2)$\times$SU(2) chiral
  model\/}, Sov. J. Nucl. Phys. {\bf 43} (1986) 657

\bibitem{Dotsenko:1990zb}
V.~Dotsenko, {\em {Solving the SU(2) conformal field theory with the Wakimoto
  free field representation}\/}, Nucl.Phys. {\bf B358} (1991) 547

\bibitem{Ooguri:1996ck} 
H.~Ooguri, Y.~Oz and Z.~Yin,
{\em D-branes on Calabi-Yau spaces and their mirrors\/},
Nucl. Phys. {\bf B477} (1996) 407, \href{http://xxx.lanl.gov/abs/hep-th/9606112}{{\ttfamily
  hep-th/9606112}}

\bibitem{Hori:2000ck}
K.~Hori, A.~Iqbal and C.~Vafa, {\em D-branes and mirror symmetry\/} (2000),
\href{http://xxx.lanl.gov/abs/hep-th/0005247}{{\ttfamily
  arXiv:hep-th/0005247}}

\bibitem{Cardy:1989ir}
J.~L. Cardy, {\em Boundary conditions, fusion rules and the {Verlinde}
  formula\/}, Nucl. Phys. {\bf B324} (1989) 581

\bibitem{Fredenhagen:2003xf}
S.~Fredenhagen, {\em Organizing boundary RG flows\/}, Nucl. Phys. {\bf B660}
  (2003) 436,  \href{http://xxx.lanl.gov/abs/hep-th/0301229}{{\ttfamily
  hep-th/0301229}}

\bibitem{Abouelsaood:1986gd}
A.~Abouelsaood, J.~Callan, Curtis~G., C.~R. Nappi, S.~A. Yost, {\em {Open
  Strings in Background Gauge Fields}\/}, Nucl. Phys. {\bf B280} (1987) 599

\bibitem{DiVecchia:1999fx}
P.~Di~Vecchia, A.~Liccardo, {\em {D-branes in string theory. 2.}\/}  (1999),
  \href{http://xxx.lanl.gov/abs/hep-th/9912275}{{\ttfamily
  arXiv:hep-th/9912275}}

\bibitem{Gaberdiel:2004nv}
M.~R. Gaberdiel, H.~Klemm, {\em {N = 2 superconformal boundary states for free
  bosons and fermions}\/}, Nucl.Phys. {\bf B693} (2004) 281,
  \href{http://xxx.lanl.gov/abs/hep-th/0404062}{{\ttfamily
  arXiv:hep-th/0404062}}

\bibitem{Gaberdiel:2011aa}
M.~R. Gaberdiel, P.~Suchanek, {\em {Limits of Minimal Models and Continuous
  Orbifolds}\/}, JHEP {\bf 1203} (2012) 104,
  \href{http://xxx.lanl.gov/abs/1112.1708}{{\ttfamily arXiv:1112.1708}}

\bibitem{Billo:2000yb}
M.~Billo, B.~Craps, F.~Roose, {\em {Orbifold boundary states from Cardy's
  condition}\/}, JHEP {\bf 0101} (2001) 038,
  \href{http://xxx.lanl.gov/abs/hep-th/0011060}{{\ttfamily
  arXiv:hep-th/0011060}}

\bibitem{Kazama:1989qp}
Y.~Kazama, H.~Suzuki, {\em New N=2 superconformal field theories and
  superstring compactification\/}, Nucl. Phys. {\bf B321} (1989) 232

\bibitem{Gaberdiel:2010pz}
M.~R. Gaberdiel, R.~Gopakumar, {\em {An AdS$_3$ Dual for Minimal Model
  CFTs}\/}, Phys.Rev. {\bf D83} (2011) 066007,
  \href{http://xxx.lanl.gov/abs/1011.2986}{{\ttfamily arXiv:1011.2986}}

\bibitem{Gaberdiel:2012uj}
M.~R. Gaberdiel, R.~Gopakumar, {\em {Minimal Model Holography}\/}  (2012),
  \href{http://xxx.lanl.gov/abs/1207.6697}{{\ttfamily arXiv:1207.6697}}

\bibitem{Creutzig:2011fe}
T.~Creutzig, Y.~Hikida, P.~B. R{\o}nne, {\em {Higher spin AdS$_3$ supergravity
  and its dual CFT}\/}, JHEP {\bf 1202} (2012) 109,
  \href{http://xxx.lanl.gov/abs/1111.2139}{{\ttfamily arXiv:1111.2139}}

\bibitem{Candu:2012jq}
C.~Candu, M.~R. Gaberdiel, {\em {Supersymmetric holography on AdS$_3$}\/}
  (2012),  \href{http://xxx.lanl.gov/abs/1203.1939}{{\ttfamily
  arXiv:1203.1939}}

\bibitem{Hanaki:2012yf}
K.~Hanaki, C.~Peng, {\em {Symmetries of Holographic Super-Minimal Models}\/}
  (2012),  \href{http://xxx.lanl.gov/abs/1203.5768}{{\ttfamily
  arXiv:1203.5768}}

\bibitem{Ahn:2012fz}
C.~Ahn, {\em {The Large N 't Hooft Limit of Kazama-Suzuki Model}\/}, JHEP {\bf
  1208} (2012) 47,  \href{http://xxx.lanl.gov/abs/1206.0054}{{\ttfamily
  arXiv:1206.0054}}

\bibitem{Candu:2012tr}
C.~Candu, M.~R. Gaberdiel, {\em {Duality in N=2 minimal model holography}\/}
  (2012),  \href{http://xxx.lanl.gov/abs/1207.6646}{{\ttfamily
  arXiv:1207.6646}}

\bibitem{Klemm:2003vn}
H.~Klemm, {\em {Embedding diagrams of the N=2 superconformal algebra under
  spectral flow}\/}, Int.J.Mod.Phys. {\bf A19} (2004) 5263,
  \href{http://xxx.lanl.gov/abs/hep-th/0306073}{{\ttfamily
  arXiv:hep-th/0306073}}

\bibitem{Kiritsis:1986rv}
E.~Kiritsis, {\em {Character formulae and the structure of the representations
  of the N=1, N=2 superconformal algebras}\/}, Int.J.Mod.Phys. {\bf A3} (1988)
  1871

\bibitem{Eholzer:1996zi}
W.~Eholzer, M.~Gaberdiel, {\em {Unitarity of rational N=2 superconformal
  theories}\/}, Commun.Math.Phys. {\bf 186} (1997) 61,
  \href{http://xxx.lanl.gov/abs/hep-th/9601163}{{\ttfamily
  arXiv:hep-th/9601163}}

\bibitem{Lerche:1989uy}
W.~Lerche, C.~Vafa, N.~P. Warner, {\em Chiral Rings in N=2 Superconformal
  Theories\/}, Nucl. Phys. {\bf B324} (1989) 427

\bibitem{Ravanini:1987yg}
F.~Ravanini, S.-K. Yang, {\em {Modular invariance in N=2 superconformal field
  theories}\/}, Phys.Lett. {\bf B195} (1987) 202

\bibitem{Reinsch:1999}
M.~W. Reinsch, J.~J. Morehead, {\em Asymptotics of Clebsch-Gordan
  Coefficients\/}, J. Math. Phys. {\bf 40} (1999) 4782,
  \href{http://xxx.lanl.gov/abs/math-ph/9906007}{{\ttfamily
  arXiv:math-ph/9906007}}

\bibitem{Ponzano:1968}
G.~Ponzano, T.~Regge, {\em {Semiclassical limit of Racah coefficients}\/},
  Spectroscopic and group theoretical methods in physics, ed.\
  F.~Bloch, Amsterdam: North-Holland Publ. Co. (1968) 1

\end{thebibliography}
\end{document}